# FEther: An Extensible Definitional Interpreter for Smart-contract Verifications in Coq

**Zheng Yang[1]*, Hang Lei[1]**

zyang.uestc@gmail.com; hlei@uestc.edu.cns

[1]School of Information and Software Engineering, University of Electronic Science and Technology of China, No.4 Section 2 North Jianshe Road, Chengdu 610054, P.R. China

*Corresponding author.

*E-mail addresses*: zyang.uestc@gmail.com (Z. Yang)

**Abstract**. Recently, blockchain technology, which adds records to a list using cryptographic links, has been widely applied in the financial field. Therefore, the security of blockchain smart contracts is among the most popular contemporary research topics. To improve the theorem-proving technology in this field, we are developing an extensible hybrid verification tool chain, denoted as FSPVM-E, for Ethereum smart contract verification. This hybrid system extends the Coq proof assistant, a formal proof-management system. Combining symbolic execution with higher-order theorem-proving, it solves consistency, automation, and reusability problems by standard theorem-proving approaches. This article completes the FSPVM-E by developing its proof engine. FSPVM-E is an extensible definitional interpreter based on our previous work FEther, which is totally developed in the Coq proof assistant. It supports almost all semantics of the Solidity programing language, and simultaneously executes multiple types of symbols. FEther also contains a set of automatic strategies that execute and verify the smart contracts in Coq with a high level of automation. The functional correctness of FEther was verified in Coq. The execution efficiency of FEther far exceeded that of the interpreters which are developed in Coq in accordance with the standard tutorial. To our knowledge, FEther is the first definitional interpreter of the Solidity language in Coq.

## 1. Introduction[1]

Blockchain technology [1], which adds records to a list using cryptographic links, is among the most popular contemporary technologies. Ethereum is a widely adopted blockchain system that implements a general-purpose, Turing-complete programing language called Solidity [2]. Ethereum enables the development of arbitrary smart contracts that can automate blockchain transactions in a virtual runtime environment, namely, the Ethereum Virtual Machine (EVM). Here smart contracts refer to the applications and scripts (i.e., programs) that execute the blockchain. The growing use of smart contracts has necessitated increased scrutiny of their security. Smart contracts can include particular properties (i.e., bugs) that expose them to deliberate attacks causing direct economic loss. Some of the largest attacks on smart contracts are well known, such as the attacks on decentralized autonomous organizations [3] and parity wallet contracts [4]. Many classes of subtle bugs, ranging from transaction-ordering dependencies to mishandled exceptions, exist in smart contracts [5]. Therefore, the security and reliability of smart-contract programs must be verified as rigorously as possible. The properties of programs can be rigorously verified by proving higher-order logic theorems. In the standard approach, a formal model for the target software system is manually abstracted using higher-order theorem-proving assistants. Such formal verification technology provides sufficient freedom and flexibility for designing formal models based on higher-order logic theories, and can abstract and express very complex systems. However, when applied to program verification, the advantages of theorem-proving technology are suppressed by automation, reusability, consistency and efficiency problems.

The above issues can be resolved by a formal symbolic process virtual machine (FSPVM) [6], which directly and symbolically executes real-world smart-contract programs using higher-order theorem-proving assistants. The program's properties are then automatically verified by the execution result. To this end, we are developing an FSPVM named FSPVM-E [7] for smart contracts deployed on the Ethereum platform. FSPVM-E is programed in Coq (a formal proof-management system) and inspired by KLEE, a high-coverage test generator for complex-systems programs [8]. Similarly to [9], the symbolic execution of FSPVM-E is verified in FEther, a hybrid proof engine that supports multiple types of symbolic execution. FEther, however, is designed for higher-order theorem proving, and its verification process is founded on Hoare [10] and reachability [11] logic. Therefore, the successful implementation of an FSPVM must overcome several challenges [6].

Some of these challenges have been addressed in our recent studies. In [6], we noted the lack of a versatile formal memory model for constructing the logic operating environment within a higher-order theorem-proving system. We thus developed a general, extensible, and reusable formal memory (GERM) framework based on higher-order logic using Coq [12]. In a later work, we extended the Curry-Howard isomorphism (CHI) [13] to resolve the basic theory of FSPVM. Herein denoted as execution-verification isomorphism (EVI), our solution combines theorem proving and symbolic execution technology. Finally, we developed an extensible large subset of the Solidity programing language, denoted as Lolisa [14], which equivalently formalizes real-world programing languages as an extensible intermediate programing language.

The present paper completes the FSPVM-E by overcoming the final challenge: developing its proof engine. Our contributions are as follows. First, we develop a definitional interpreter in Coq's specification language (Gallina). This interpreter symbolically executes the smart contracts of Ethereum written in Lolisa on the GERM framework. The execution results are represented by a GERM logic memory state, which can be verified in Coq. Next, we implement a set of automatic evaluation strategies based on the Ltac [12] mechanism, by which FEther finishes the execution and verification process. The correctness of FEther is then certified in Coq. The present FEther is the optimized version with higher evaluation efficiency than the interpreters developed in Coq using standard tutorial approaches. To our knowledge, FEther is the first hybrid proof engine specification that automatically and symbolically executes and verifies Ethereum smart contracts in Coq.

The remainder of this paper is structured as follows. Section 2 describes the difference between the FEther and other relevant works. Section 3 introduces the foundations of the present work, including the prototype, the basic environment of Lolisa, and the preparatory modification of the GERM framework. Section 4 describes the theoretical design and implementation of FEther, and its self-correctness certification. Section 5 verifies FEther in a real-world case study and analyzes its benefits. Section 6 discusses the advantages and limitations of FEther. The study concludes with Section 7.

[1] Abbreviations: EVM, Ethereum Virtual Machine; FSPVM, formal symbolic process virtual machine; FSPVM-E, formal symbolic process virtual machine for Ethereum; CHI, Curry-Howard isomorphism; EVI, execution-verification isomorphism; FSPVM, formal symbolic process virtual machine; API, application programming interface; GADT, generalized algebraic datatype; GERM, general, extensible, and reusable formal memory; BMC, bounded model checking; CBNT, call-by-name termination; IRE, information redundancy explosion; CR, concurrent reduction; TCB, trusted computation base; TCOC, trusted core of Coq; ISA, instruction set architecture.

## 2. Related Works

The security of smart contracts has been seriously researched since 2015. The security of smart contracts and similar lightweight programs can be rigorously guaranteed by formal methods. Our symbolic executor has several novel features that distinguish it from other approaches. This section introduces the interesting achievements already reported in this field.

The EVM execution is formally described in Yellow Paper [15]. This official document also provides the data, algorithms, and parameters required for building consensus-compatible EVM clients and Ethereum implementations. Yellow Paper, however, does not always clarify the operational behavior of the EVM. In such cases, it is often easier to consult an executable implementation for guidance.

Most of the recent researches have concentrated on EVM security. The C++ implementation Cpp-ethereum plays a dual role of security and defector semantics in EVMs. Lem semantics [16] is a Lem [17] implementation of EVM providing executable semantics of EVM, which formally verifies smart contracts. However, the Lem semantics do not precisely capture the inter-contract execution. KEVM [18] is a formal semantics for EVMs, resembling Lem but written in the K-framework. As KEVM is executable, it can run the validation test suite provided by the Ethereum foundation. According to the authors of [18], the KEVM reference interpreter passes the full 40,683-test EVM compliance suite. Nevertheless, self-correctness cannot be proven completely or even certified in KEVM. Moreover, none of the above approaches satisfies the de Bruijn criterion [19].

Mythril [20] is a security analysis tool for Ethereum smart contracts. Mythril detects various problems by concolic analysis, but whether the tool effectively increases the reliability of smart contracts has not been proven.

The above researches adopt the bytecode of Solidity. The consistency between Solidity and bytecode after compiling cannot be guaranteed. However, high-level formal specifications and relevant formal verification tools of Solidity, which are important for programing and debugging smart contract software, have received little attention.

Finally, some of these works focus on a specific domain. Their complex architecture is inflexible and not easily extendible to new relevant problems.

## 3 Foundational Concepts

The present paper builds upon our recent previous works. Therefore, prior to defining the formal specifications of FEther, we first define the basic environment.

### 3.1 Predefinitions

Previously, we constructed a prototype of our FSPVM framework, which extends the proof assistants as a hybrid system combining symbolic execution and higher-order theorem proving. This prototype solves consistency, automation, and reusability problems in standard theorem-proving approaches. The prototype, which is the blueprint of FSPVM-E, consists of the following two parts.

The first step combines symbolic execution and higher-order theorem-proving, and simulates the execution of real-world programs. The prototype is verified by solving the consistency, automation, and reusability problems of higher-order theorem-proving. The blueprint (denoted as FSPVM) extends the higher-order theorem-proving assistants that support CHI. Specifically, we extend CHI to EVI, which operates under Principles 1 and 2 below:

$$\textit{verifications correspond to proofs} \quad (1).$$

$$\textit{verifications correspond to execution of programs} \quad (2).$$

FSPVM also contains a formal general memory model $\mathcal{FM}$ and a formal intermediate language $\mathcal{FL}$ (equivalent to the respective general-purpose programing language $\mathcal{L}$), for rewriting the formal versions of *RWprogram* as *FRWprogram*. It also requires a formally verified interpreter (FInterpreter), which plays the roles of the execution-engine core in the FSPVM (i.e., simulating the real execution process of *FRWprograms* and generating the logic memory states; see Figure 1). The executable semantics of $\mathcal{FL}$ provide the instruction set architecture (ISA) of FInterpreter. Currently, we are developing a specified extensible verified FSPVM in Coq for Ethereum smart-contracts verification, denoted as FSPVM-E. In FSPVM-E, the $\mathcal{FM}$ and $\mathcal{FL}$ are the GERM framework and the Lolisa programing language, respectively.

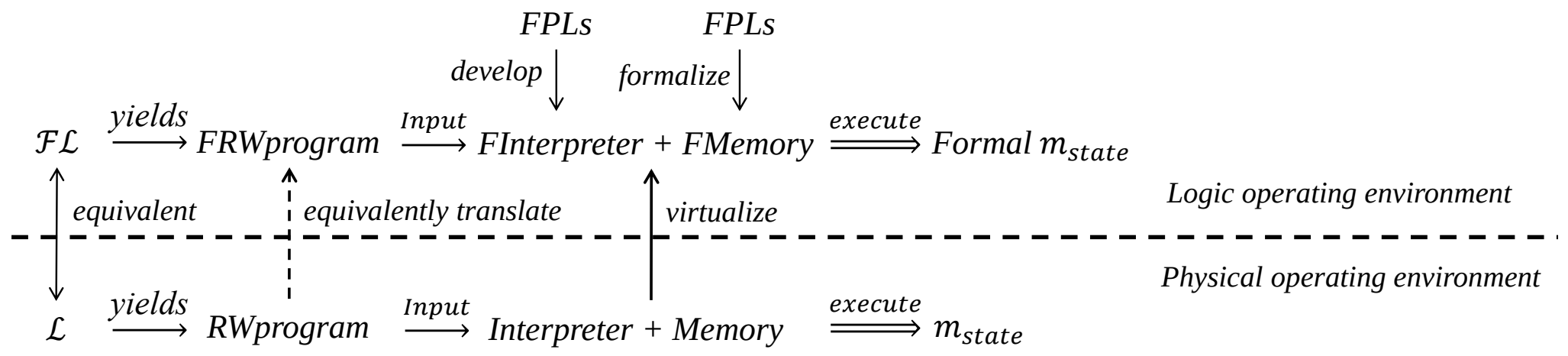

Figure 1. Equivalence of real world program (*RWprogram*) execution and execution in a logic environment

***GERM*** The GERM is a general, extensible, and reusable formal memory framework. It simultaneously supports different formal verification specifications, particularly at the code level. This framework simulates the physical memory hardware structure, including a low-level formal memory space, and provides a set of simple, nonintrusive application programing interfaces (APIs). The proposed GERM framework is independent and customizable, and was verified entirely in Coq. Table 1 summarizes the top-level interface of FEther, where $\mathrm{m_{value}}$ and $a$ represent a memory value of type *value* and a memory address of type $L_{address}$, respectively. In the specific formal specification, a formal memory state has type *memory*. Finally, $b_{infor}$ represents the block information for environment checking.

***Lolisa*** The $\mathcal{FL}$ is Lolisa, a large subset of the Solidity programing language. Assisted by generalized algebraic datatypes (GADTs) [21], the formal syntax of Lolisa adopts a stronger static-type system than Solidity, which enhances the type safety. Lolisa includes contract declarations (*Contract*), modifier declarations (*Modifier*), variable declarations (*Var*), structure declarations (*Struct*), assignments (*Assign*), returns (*Return*), multi-value returns (*Returns*), throws (*Throw*), skips (*Snil*), function definitions (*Fun*), while loops ($Loop_{while}$), for loops ($Loop_{for}$), function calls ($Fun_{call}$), and conditional (*If*) and sequence statements.

| Function | Description |
|---|---|
| $read_{dir}$ | Read $\mathrm{m_{value}}$ from $a$ directly |
| $read_{chck}$ | Read $\mathrm{m_{value}}$ from $a$ after validation checking |
| $write_{dir}$ | Write $\mathrm{m_{value}}$ at $a$ directly |
| $write_{chck}$ | Write $\mathrm{m_{value}}$ at $a$ after validation checking |
| $address_{offset}$ | Offset address $a$ to $a'$ |
| $allocate$ | Allocate memory blocks |
| $free_{mem}$ | Free a specified memory block |
| $init_{mem}$ | Initialize the entire memory space |

Table 1. Basic memory-management APIs employed in the formal memory model.

Table 2 summarizes the helper states used in the dynamic semantic definitions, and Table 3 lists the helper functions for calculating commonly needed values in the current program state. All of these state functions will be encountered in the following discussion. The components of specific states will be denoted by appropriate Greek letters subscripted by the state of interest. In Table 2, $M$, $\sigma$, and $\mathcal{E}$ denote the contexts of the formal memory space, a specific memory state, and the execution environment, respectively. The proof evaluation is executed in the proof contexts, denoted as $\Gamma$, $\Gamma_1$, … For brevity, we hereafter represent the overall formal system by $\mathcal{F}$, the current execution environment by $env$, and the super-environment of type *env* by $fenv$.

In the following sections, we introduce the relevant analysis and solutions that improve the computation efficiency of $\mathcal{FI}$ in higher-order theorem-proving assistants.

| | | | |
|---|---|---|---|
| $\mathcal{E}$ | environment information | $\mathcal{F}$ | formal system world |
| $M$ | memory space | $\Gamma$ | proof context |

Table 2. State functions in the dynamic semantic definitions

To simplify the verification process and the development of the respective formal verified Lolisa interpreter in Coq, we maintained the Lolisa programs as structural programs. For this purpose, the semantics of Lolisa were forced to adhere to the following pointer counter axiom. The conventions of the *Struct* datatypes are defined in Convention 1. The Pointer Counter axiom is a FEther design principle that maintains Lolisa as a structural language.

**Axiom** (Pointer Counter): If statement $s$ is the next execution statement, it must be the head of the statement sequence in the next iteration.

| | | | |
|---|---|---|---|
| $set_{env}$ | Changes the current environment | $env_{check}$ | Validates the current environment |
| $id_{search}$ | Searches the address of array elements | $init_{var}$ | Initializes variable into memory block |
| $id_{map}$ | Searches the address of mapping elements | $eval_{str}$ | Evaluates the struct datatype |
| $eval_{bop}$ | Evaluates binary expression of Lolisa value | $eval_{uop}$ | Evaluates unary expression of Lolisa value |
| $mems_{find}$ | Searches the valid struct field | $eval_{head}$ | Evaluates the struct value basic information |

Table 3. Helper functions

*Type level* ⌣ *Declare type annotation*
*Value level* ⌣ *Represent struct value & field value*
*Expression level* ⌣ *Represent right value of struct type*
*Statement level* ⌣ *Declare new struct type*

Convention 1. Formal *Struct* datatypes

To avoid infinite loops in the programs, FSPVM also imports bounded model checking (BMC) [22]. Fortunately, the EVM does not support infinite execution processes, as each execution step costs the *gas* of the smart contract owners. If the *gas* balance cannot satisfy the limitation, the execution terminates. This design well suits the BMC concept. Therefore, our implementation uses *gas* to limit the execution of the Solidity programs.

In the following contents, we represent other arguments by the wildcard " * " and by the symbol $\{| * \longmapsto * |\}.E$ is the syntax of constructor pattern matching of the $\lambda$-expression $E$ [23]. To avoid ambiguity in the following discussion of FEther, the functions represent the programs and functions written in Gallina, and *RWprogram* represents the real-world programs written in general-purpose programing languages.

### 3.2 Modifications for optimization

As mentioned previously, when analyzing the current problems, the computational efficiency of the definitional interpreter based on the FSPVM may be extremely low. The three essential problems are *call-by-name termination* (CBNT), *information redundancy explosion* (IRE), and *concurrent reduction* (CR). To optimize the low-level computations of the evaluation problems, we incorporate the respective solutions in [25] into the implementation details of FEther.

First, the sequence statement $s$ is implicitly replaced by an equivalent list rather than explicitly defined, which avoids the CBNT problem. Second, the pattern matchings and reusable functions are encapsulated as optimization helper functions. Some of these functions are summarized in Table 4. To avoid the CR problem, we finally impose a limitation $K$ (independent of the *gas* constraint) on the expression and value layers.

| | | | |
|---|---|---|---|
| $set_{env}$ | Changes the current environment | $env_{check}$ | Validates the current environment |
| $id_{search}$ | Searches the address of array elements | $init_{var}$ | Initializes variable into memory block |
| $id_{map}$ | Searches the address of mapping elements | $eval_{str}$ | Evaluates the struct datatype |
| $eval_{bop}$ | Evaluates binary expression of Lolisa values | $eval_{uop}$ | Evaluates unary expression of Lolisa value |
| $mems_{find}$ | Searches the valid struct fields | $eval_{head}$ | Evaluates the struct value basic information |
| $lexpr_{check}$ | Checks the validation of the l-position expressions | $get_{bool}$ | Gets the Boolean value from memory value |
| $valid_{array}$ | Check the declaration of array type | $get_{stt}$ | Gets the statement value from memory value |

Table 4. Encapsulation functions in the optimized FEther

## 4. FEther Implementation

FEther is the bridge that connects the GERM framework, the Lolisa programing language, and the trusted core of Coq (TCOC). As demonstrated in our previous work and elaborated in the following subsections, FEther can be totally built in Coq.

### 4.1. Architecture

Figure 2 shows the overall structure of the FEther framework. The whole FEther is constructed in the trusted domain of Coq, and logically comprises three main components: a parser, an ISA based on Lolisa semantics, and a validation checking mechanism (see left, center and right blocks in Figure. 2, respectively). The parser analyzes the syntax of the *FRWprograms* written in Lolisa. According to EVI theory, FEther is essentially a huge function written in Gallina. In this sense, it differs from the real-world virtual machines of high-level programing languages such as Smalltalk, Java, and .Net, which support bytecode as their ISA and are implemented by translating the bytecodes of commonly used code paths into native machine code. Instead, the ISA of FEther comprises the Lolisa semantics, which specify the semantics of the syntax tokens that govern the respective behaviors. The validation checking mechanism includes two parts: checking the result validation (including the memory states and values), and checking the execution condition. First, because all functions are vulnerable to undefined situations, they are developed with the help of effect programing. More specifically, all functions are tagged by an optional type. A valid result is returned in the form *Some t*; an invalid result is returned as an undefined value *None*. The symbol $[\![t]\!]$ denotes that term *t* is tagged by an optional type. In the second part, the *gas* and *K* limitations are validated by the helper functions $env_{check}$ and $pump_{check}$, respectively.

FEther inherits the low-coupling property from Lolisa. Within the same level, the executable semantics are wholly independent, and are encapsulated as modules connected by a set of interfaces. In different levels, the higher-level semantics can access the lower-level semantics only via the interfaces, and the implementation details of the lower-level semantics are transparent to the higher-level semantics (indicated by dotted lines in Figure 2). Moreover, the implementation of the higher-level semantics does not depend on the lower-level semantics.

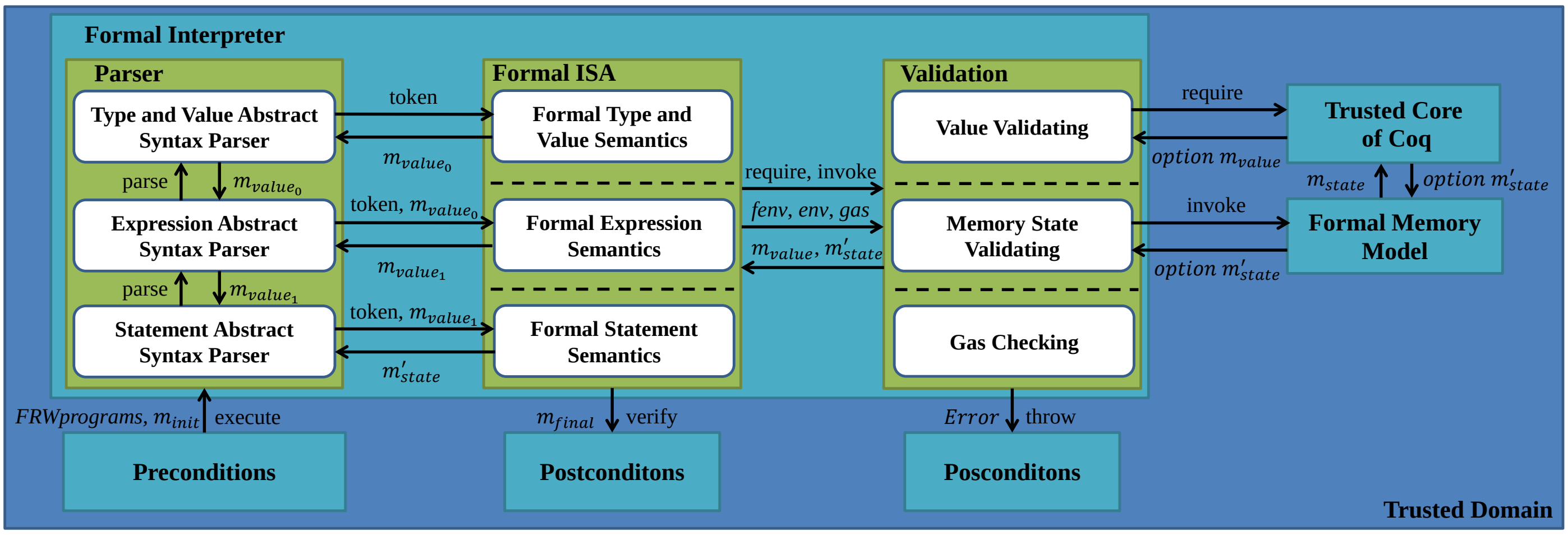

Figure 2. Architecture of FEther

The workflow of FEther is outlined in Figure 3. The user first sets the initial memory state and the target *FRWprograms* by initializing the formal memory space of GERM and applying the translator. Note that the translator is an optional auxiliary component, which translates the Solidity programs into the *FRWprograms* written in Lolisa. To this end, it searches the abstract syntax tree of Lolisa, binds the variable identifiers with a unique memory address, and declares the ML modules. This process can also be completed manually. Because the translator is part of FSPVM-E rather than a core component of FEther, it is not further discussed in this paper. Next, the FEther parser analyzes the *FRWprograms* according to the Lolisa abstract syntax tree, and invokes the respective executable semantics. The TCOC handles the evaluation requirements, and the results are validated by the validation mechanism. Although the validation module is logically independent of the other parts (as mentioned above), it is implemented separately in the *Formal Interpreter* and *Formal Semantic* modules in real cases. Therefore, the validation module is not explicitly defined in Figure 3. The final formal memory state will be assumed in the property theorems.

Lolisa is defined by GADT, which guarantees well-formed constructions of the syntax specifications. Thus, the side conditions of syntax correctness do not need checking by help functions defined in FEther. The type safety can be checked by the type-checking mechanism of Coq. The complete workload of constructing an FEther framework with 100 memory blocks is itemized in Table 5.

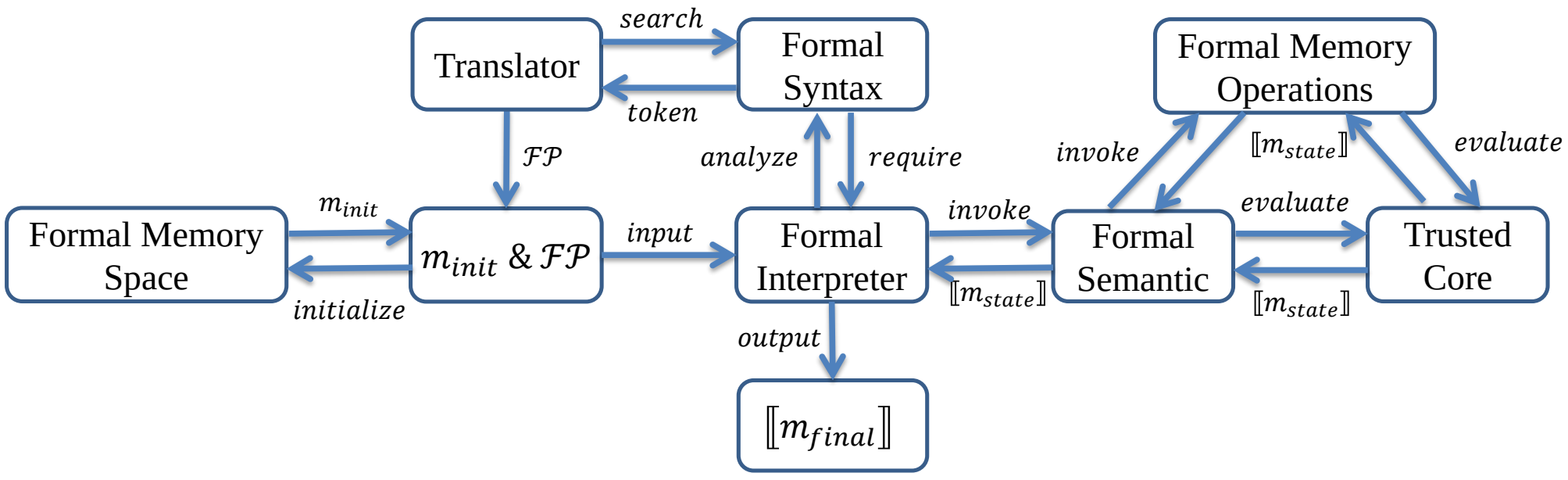

Figure 3. Workflow of FEther

| | Objects | Lines in Coq |
|---|---|---|
| Formal Type and Value layer | 58 | 1,347 |
| Formal Expression layer | 46 | 1,503 |
| Formal Statement layer | 40 | 741 |
| Automatic tactic | 26 | 344 |
| Correctness Lemmas | 74 | 3,746 |
| Total | 244 | 7,681 |

Table 5. Workload statistics for constructing the FEther framework with 100 memory blocks.

### 4.2 ISA of FEther

The ISA of FEther is the core of the proof engine, which follows the big-step operational semantics of Lolisa. As shown in Figure 2, the FEther ISA is separated into three layers. FEther is implemented as described in Appendix A.

#### 4.2.1 Value layer

This project aims to formalize a mechanized syntax and semantics for a subset of the Solidity language, which can be directly executed and verified in higher-order logic theorem-proving assistants. Therefore, the Solidity values must be evaluated like the native values in the formal system. Ideally, the values of Solidity or some mainstream high-level programing language would be explicitly employed in the formal system. Due to the strict typing system of the trusted core and the adoption of different paradigms, however, Gallina (Coq) does not directly support array, mapping, and other complex values. Therefore, we must define an interlayer between the values of the real-world language and the native values of the formal system. This interlayer directly represents the real world-values by an equivalent syntax, and translates them into the native values using formal semantics.

After evaluating the Lolisa value by formal executable semantics, the native value information is computed or derived in the base formal system, and the respective GERM memory values are determined. In the following sections, $\mathcal{ESV}$ represents the entry point of calling the value semantics, which is abstractedly defined by Rule 3 below.

$$\mathcal{ESV} :: \mathbb{Z} \rightarrow (\forall\, \tau{:}\, type, val\ \tau) \rightarrow memory \rightarrow Blc \rightarrow Env \rightarrow option\ value. \quad (3)$$

Here, the metavariable $val$ incorporates the Lolisa value $val$ and the mapping value $val_{map}$. Apart from lacking definitions related to mapping type, $val_{map}$ has the same static typing rule as $val$, so the two values can be combined. The mapping relation between each Lolisa value and its unique memory value is expressed as $\approx$. The memory state $\sigma$ remains unchanged after each value evaluation. Following the definition order given in [14], we then define the computational semantics of the Lolisa value.

First, we define the computational semantics of the constant values. In Lolisa, the constant values are the set of normal-form values, and the set of metavariables $v_{const}$. The $v_{const}$ evaluation process generates the respective memory values for directly recording the native value information of the formal system. For example, consider the constant variable $Vbool(b){:}\, Tbool \in v_{const}$. The computational semantics of $Vbool(b)$ are defined by Rule 4. We then define $\mathcal{ESV}_{const_{bool}}(n, env, b_{infor}) \approx Vbool(n)$.

$$\mathcal{ESV}_{const_{bool}} \equiv \lambda\ (n{:}\, bool).\, \lambda\ (env{:}\, Env).\, \lambda\ (b_{infor}{:}\, Blc).\, Vbool(n) \Rightarrow Some\ Bool(n, env, b_{infor}). \quad (4)$$

The computational semantics $\mathcal{ESV}_{const}$ are summarized in Table A.1 of A. The correctness of $\mathcal{ESV}_{const}$ is certified by Theorem 1 (the constant-mapping theorem).

**Theorem 1** (Constant mapping) For all Lolisa values $v_{const}(n)$, environment values *env*, and block information $b_{infor}$, the mapping $\mathcal{ES}_{const}(n, env, b_{infor}) \approx v_{const}(n)$ holds.

We then define the semantics of the reference values (the array, mapping, structure and field access values), which are needed for accessing the formal memory space and match indexes. The respective values are defined as follows, and the semantics $\mathcal{ESV}_{array}$ of the array values are defined in Table A.2 of A. Here, $id_{search}$ is a subsidiary function. Because the Lolisa language supports an *n*-dimensional array (by Rule 6), $id_{search}$ is a special subsidiary function that searches the memory block indexed by the current *n*-dimensional array index. $id_{search}$ is also used in the $init_{var}$ function. Below we introduce the specific implementations of $id_{search}$. The abstract function of $id_{search}$ implementations is given by Rule 5.

$$id_{search} :: type \rightarrow L_{address} \rightarrow memory \rightarrow Env \rightarrow option\ address \quad (5)$$

$$Tarray\ \Big[id_0\ Tarray\ \Big[id_1\ Tarray\ \big[\ldots\ [Tarray\ id_n\ \tau_{final}]\big]\Big]\Big] \quad (6)$$

The mapping value is stored as a singly linked list structure in the GERM memory model, and in the form

$$Map{:}\, L_{address} \rightarrow option\ (prod\ value_{map}\ value) \rightarrow type_{map} \rightarrow type \rightarrow option\ address \rightarrow Env \rightarrow Blc \rightarrow value$$

in Lolisa. In the above expression, the first parameter stores the initial address, the second parameter stores the paired key and indexed values, the third and fourth parameters record the key value and indexed value types, respectively, and the fifth parameter represents the next address. The *Map* can be briefly abstracted as Figure 4. In this design, the structure supports the *n*-dimensional mapping datatype. For instance, consider the 2-dimensional mapping datatype $mapping\ \big[\tau_{map_0} \Rightarrow mapping\ [\tau_{map_1} \Rightarrow \tau_{final}]\big]$ in Figure 5. Each memory block in one dimension is the initial block of a respective mapping list in two dimensions. The *n*-dimensional mapping type in Lolisa can be defined by the same process. Obviously, the $id_{map}$ function can be implemented by any singly linked list-search algorithm. If the search is successful, the function returns $Some\ addr$; otherwise, it returns $None$ (see Table A.3 in A).

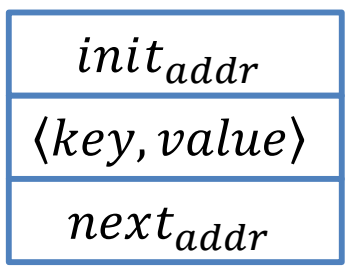

Figure 4. Abstract structure of a mapping-type memory block

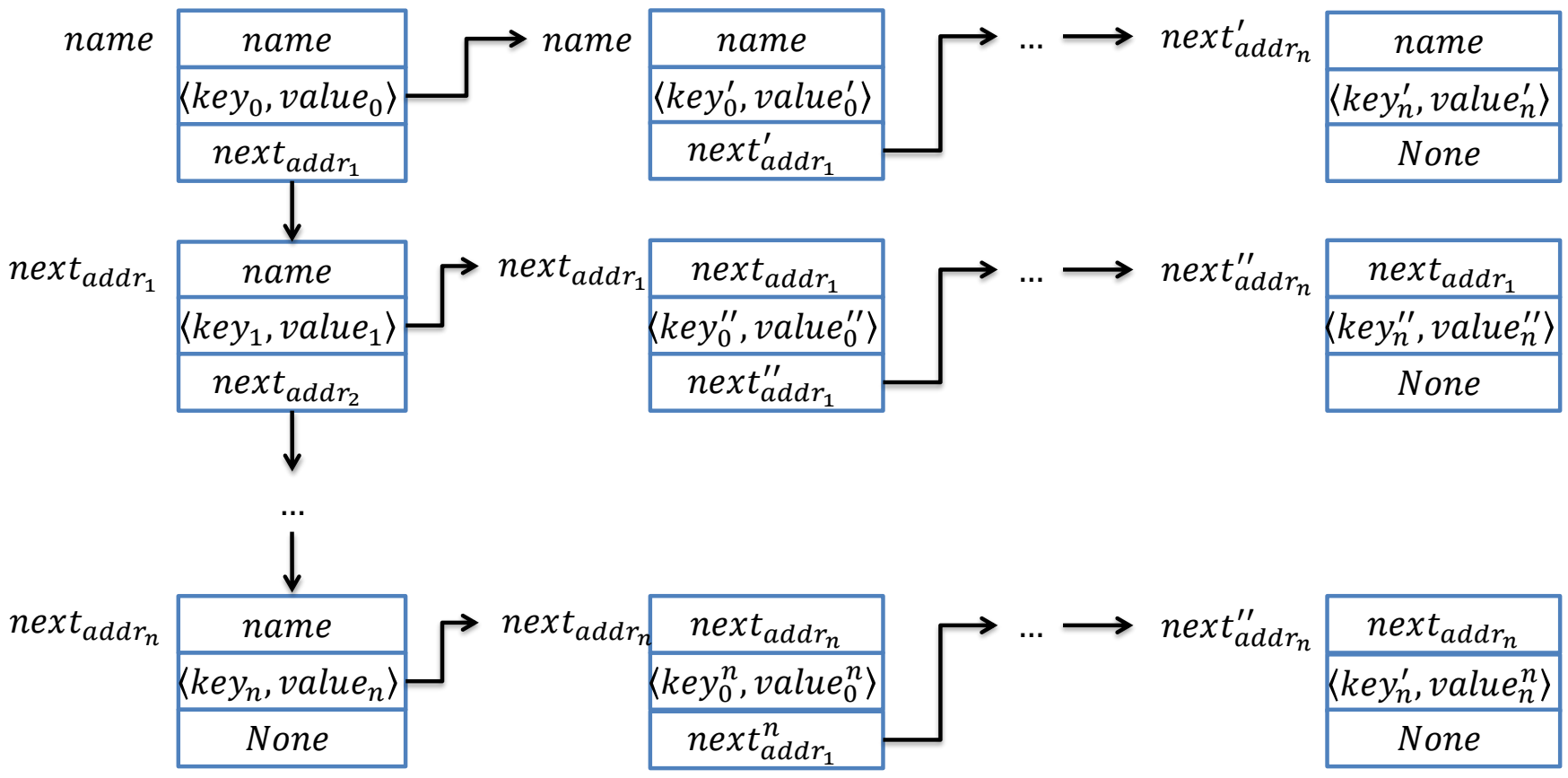

Figure 5. Structure of a 2-dimensional mapping value stored in GERM

At the value level, a *Struct* memory value is represented by a struct datatype. Therefore, it resembles a normal-form value, and can be extracted directly by $read_{check}$ (see Table A.4 of A).

The Solidity semantics of field access are very complex and consist of two parts: a contract member access and a struct field access. If the contract member access derives from an inheritance relationship or a special identifier, such as the keyword "*this*", it can be accessed directly through the ML module system. Briefly, the module scope of each function and contract is defined by the translator, and can be accessed by the ML module system provided by Coq. If the contract member access derives from a variable, the contract information stored in the respective memory block is searched, and the identified member is accessed by the field access mechanism of the Coq built-in ML module system. The struct field access semantics are supported by Lolisa. The built-in EVM functions in a standard structure, such as *msg* or *block*, are of no concern because they are already defined in the Lolisa standard library [14]. Therefore, they can be treated as normal structures in the semantics. We denote $eval_{head}$ as the process of evaluating a base address $a_{init}$ and a struct-type address $a_{type}$. Taking $mem_0.mems.mem_n$, $a_{init}$, and $a_{type}$ as parameters, $mems_{find}$ then seeks the memory value indexed by $mem_n$. In this process $mems_{find}$ repeatedly applies $read_{chck}$ to acquire the list of member values. If successfully invoked, $mems_{find}$ returns a pair $(Dad, m_v)$. Here, $Dad$ refers to the address of $mem_{n-1}$ because $mem_{n-1}$ is an implicit argument to Solidity when $mem_n$ is a function-call pointer. A common usage in Solidity is $a.send(v, mss) \equiv send(a, v, mss)$, which indicates that the identifier $a$ is a parameter of the $send$ function during interpretation or compilation. Therefore, if $m_v$ is a pointer to a function call $\big(Fid(*)\big)$, $Dad$ and the function input $opars$ should be combined into $m_v$ to facilitate their transmission to the next level. The above evaluation process is summarized in Table A.5 of A.

### 4.2.2 Expression Layer

The executable semantics of expressions are the rules that acquire the results of the value layer and evaluate the Lolisa expression in the memory value of GERM. The evaluation requires the left-value (l-value) and right-value positions, representing the memory addresses and the specific memory value, respectively. In the following contents, the entered pointer of the expression layer $\mathcal{ESE}$ is defined by Rules 7 and 8.

$$\mathcal{ESE}_l :: \mathbb{Z} \to (\forall\, \tau_0\ \tau_1 \colon type, expr\ \tau_0\ \tau_1) \to memory \to Blc \to Env \to option\ L_{address}, \quad (7)$$

$$\mathcal{ESE}_r :: \mathbb{Z} \to (\forall\, \tau_0\ \tau_1 \colon type, expr\ \tau_0\ \tau_1) \to memory \to Blc \to Env \to option\ value. \quad (8)$$

In formal Lolisa semantics, the modifier expression is a special one that cannot be evaluated in the expression layer (as will be later explained in the statement semantics). The computational semantics are defined as follows:

$$\mathcal{ESE}_{lmodi} \equiv \mathcal{ESE}_{rmodi} \equiv\ Error.$$

***Expressions in the l-value position:*** The following rules define the semantics of evaluating the expressions in the l-value position (i.e., the respective memory address). The expressions in the l-value position, which can be constructed by the *Econst* constructor, represent Lolisa values at the expression level. Specifically, the left values can be assigned as the *Econst* specified by *Varray* and *Vmap*. As mentioned previously, *Varray* and *Vmap* are address pointers to values stored in specific memory blocks. This construct (shown in Example 1) is commonly used in most general-purpose programing languages.

$$A[i] = a;\ \text{(Example I).}$$

Thus, *Varray* and *Vmap* can represent not only memory values, but also memory addresses. Note that the remaining values (*Vstruct* and *Vfield*) are also address pointers by specifying the *Econst* constructor, but cannot represent the expressions in the l-value position. Because *Evar* can represent any variables address using any types, including structure and field access values. To avoid confusion between *Evar* and *Econst* specified by *Vstruct* and *Vfield*, we set the convention that *Vstruct* and *Vfield* represents only the memory value at the value level. In both Solidity and Lolisa, *Vfield* admits many special structures, such as *msg* and *block*, whose members cannot be changed at will. In rare cases, Solidity allows field-access expressions in the left position. Therefore, to ensure that Lolisa remains well-formed and well-behaved, the left value in expressions cannot be evaluated by *Vfield*. The fields of structures can be changed by invoking *Estruct* to change all fields, or by declaring a new field, as explained below. Although the limitations of *Vstruct* and *Vfield* are inconvenient for programmers and verifiers, they avoid any potential risk.

Therefore, if the constructor is *Varray*, the semantics are chosen as $\mathcal{ESE}_{lexpr_{array}}$ (defined in Table A.6 of A); if the constructor is *Vmap*, they are chosen as $\mathcal{ESE}_{lexpr_{map}}$ (Table A.7 of A). The other constructors return *None* directly. The semantics of the left constant value $\mathcal{ESE}_{lexpr_{const}}$ at the expression level are then given in Table A.8 of A. The IRE problem is avoided by introducing a special helper function $lexpr_{check}$, which encapsulates the matching tree for obtaining the value information recorded in valid constructors.

The reference expressions *Evar*, *Efun*, *Econ*, and *Epar* need only to return their addresses directly. In Table A.9 of A, these reference expressions are summarized as $Eaddr(\llbracket name \rrbracket){:}\,expr\ eaddr(\llbracket name \rrbracket)\ eaddr(\llbracket name \rrbracket)$. The *Estruct*, *Ebop*, and *Euop* expressions can only be assigned as right expression values, so their semantics are banned by the $lexpr_{check}$ function (which returns an undefined result *None*).

***Expressions in the r-value position:*** The following functions describe the semantics of evaluating expressions in the r-value position (i.e., the respective memory values). The evaluation of constant expressions is given in Table A.10 of A. Assisted by $\mathcal{ESV}_{value}$, the $\mathcal{ESE}_{rexpr_{const}}$ directly provides the respective memory value.

According to Convention 1, the struct constructor *Estruct* represents an expression value at the right position, which is the only way to initialize or modify struct-type terms. The semantics of the right struct value are defined in Table A.11 of A. The helper function $eval_{str}$ contains the type matching and value evaluation. The type matching part checks whether the type of each value satisfies the respective field. The second part recursively invokes $\mathcal{ESV}_{value}$ to evaluate the values in the respective memory values. If the evaluation process yields a *None* message, the *Estruct* evaluation has failed. Otherwise, the members' value set is retrieved and the respective struct memory value is returned.

The semantics of the address-pointer expressions are defined in Table A.12 of A. As shown in the formal semantic definitions, the results are obtained by applying $read_{chck}$ directly.

Finally, the semantics of the binary and unary operations are defined in Tables A.13 and A.14 of A, respectively. Because of the static-type limitation in the formal abstract syntax definition based on GADTs, all expressions, sub-expressions and operations are well-formed, and the semantics do not need to check the type dependence relation. Therefore, subsidiary assist functions are not required. The functions $eval_{bop}$ and $eval_{uop}$ take the results of the expression evaluations and the required operations as arguments, and combine them to generate new memory values. In the present version of FEther, the above definition forbids mixed arithmetic operations, such as "int + float," because Solidity does not completely support the float datatype, and float values are rarely employed in smart contract programs. Therefore, mixed-arithmetic operations would add unnecessary complexity and computational burden when implementing the formal interpreter.

### 4.2.3 Statement Layer

Having defined the semantics, we can now define the statement layer. Statement semantics parse the *FRWprograms* written by Lolisa, and evaluate the new memory states. The semantics of the sequence statements are not explicitly defined, and the relevant statement definitions are modified to improve the extremely low computational efficiency of solving the CBNT problem. We express the evaluation process of a statement as $\mathcal{ESS}$, and give its abstract definition as Rule 9.

$$\mathcal{ESS} :: \mathbb{Z} \rightarrow memory \rightarrow option\ (list\ value) \rightarrow Env \rightarrow Env \rightarrow statement \rightarrow option\ memory. \qquad (9)$$

Most statement evaluations employ the helper function $env_{check}$, which takes the current environment *env* and the super-environment *fenv* as arguments, and checks the conditions (gas limitation and execution-level validity). For example, if the domains in *env* and *fenv* are equal but have different execution levels, the program is terminated and *env* is reset by *fenv*. If $env_{check}$ returns a *true* result, the current statements are executed; otherwise, the program is terminated and the initial memory state is restored.

Contract declarations are among the most important Solidity statements. Contract declaration in Lolisa involves two operations. First, the consistency of the inheritance information is checked using the helper function $inherit_{check}$, which determines whether the current inheritance relation $inherits$ is stored in the current module context $\mathcal{C}$. The function $inherit_{check}$ is defined as a sum type in Rule 10:

$$inherit_{check} \equiv \forall\,(inherits\ inhertis_c: list\ address), \{inherits = inhertis_c\} + \{inherits \neq inhertis_c\}. \tag{10}$$

Second, the initial contract information, including all member identifiers, is written into a designated memory block by the assistant function $write_{dir}$. The formal semantics of the contract declaration are defined in Table A.15 of A.

Variable declaration is a basic task in Lolisa. The function $init_{var}$ is a special case of $write_{dir}$, with type given by Rule 11.

$$init_{var} :: memory \to Env \to Blc \to option\ access \to type \to address \to option\ memory. \tag{11}$$

This function takes the current memory state, variable type, indexed address, and environment information as parameters, and initializes the respective memory block. Being based on the GERM memory model, the initialization and location processes of this term with the array datatype differ from those in standard researches on formalizing array types. The function $init_{var}$ calls the $init_{array}$ function to initialize the respective terms.

In Yang et al. [14], *normal types* are datatypes whose typing rules disallow recursive definition. A normal type is assigned as $\tau_{final}$. The type $\tau_{final}$ is the recursive base of a multidimensional array. In other words, if the $\tau$ of the current element is $\tau_{final}$, it represents the final dimension of the recursive definition of the current multidimensional array.

Particularly, because each memory block in the GERM memory model directly stores all logic information with type *value* [6], regardless of the sizes of the array elements, we only need to calculate the number of logic elements in the array. For example, consider the following three-dimensional array $a[2][3][2]$ with type

$$Tarray\ \Big[iAconst_id(2)\ Tarray\ \Big[iAconst_id(3)\ Tarray\ \big[iAconst_id(2)\ \tau_{final}\big]\Big]\Big].$$

The full tree structure of this array is given in Figure 6, and the mathematical evaluation process is shown as follows:

$$\begin{aligned}&\big(size_1 + size_1 * (size_2 + size_2 * size_3)\big)\\ &= \big(2 + 2 * (3 + 3 * 2)\big)\\ &= 20.\end{aligned}$$

. Note that this array requires a memory allocation of 20 blocks.

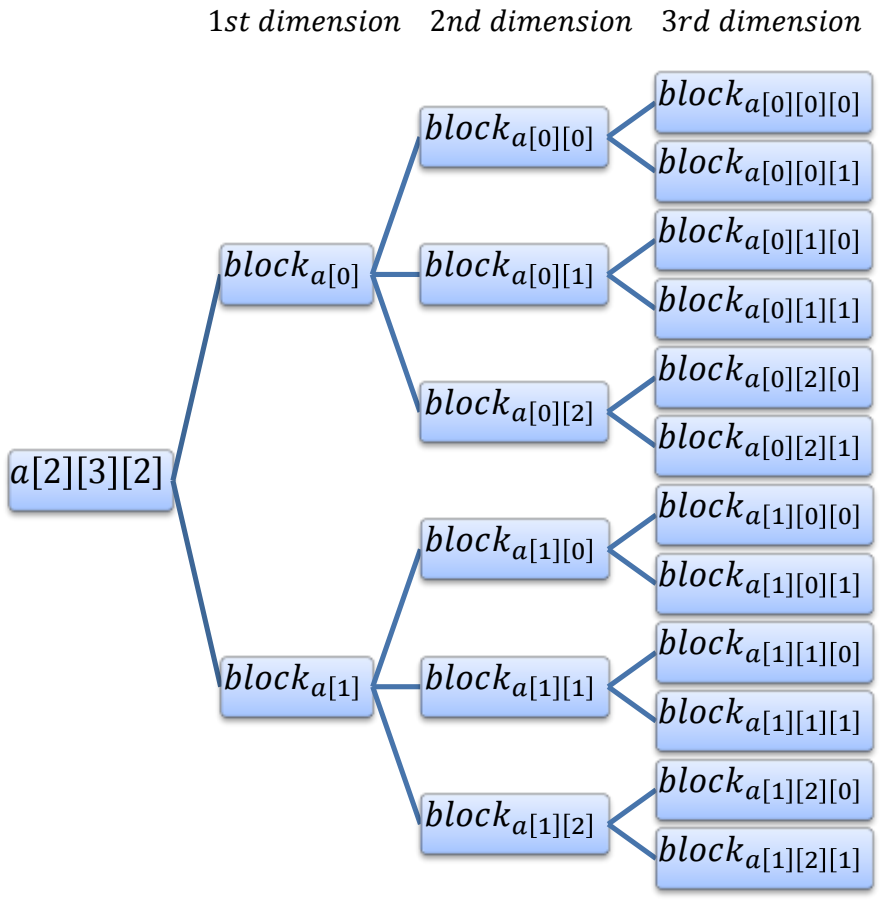

Figure 6. Example of a 3-dimensional array

The array-size calculation can be summarized as Rule 13. Using this formula, we can implement the subsidiary function $id_{array}$ to calculate and return the number of blocks allocated to arrays in each dimension. The abstract of this assignment is defined in Rule 14. Note that $size_1$ in Rules 12 and 13 represents the size of the current dimension rather than the size of the first dimension. For example, to calculate the $array_{size}$ of the second dimension of $a[2][3][2]$, we should replace $size_1' = 2$ in Rules 12 and 13 with $size_2' = 3$.

$$array_{size} \equiv size_1 + size_1 * \left(size_2 + size_2 * \left(\ldots\left(size_{n-1} + size_{n-1} * size_n\right)\right)\right) \equiv \sum_{i=1}^{n} \prod_{j=1}^{i} size_j. \tag{12}$$

$$group_{size} \equiv array_{size} \; / \; size_1 \equiv \left(\sum_{i=1}^{n} \prod_{j=1}^{i} size_j\right) / \; size_1. \tag{13}$$

$$id_{array} :: index_{array} \rightarrow memory \rightarrow Env \rightarrow option\ \mathbb{Z}. \tag{14}$$

Figure 7 shows the initialization process of $a[2][3][2]$, which follows its tree structure. In step (1), FEther searches a continuous memory space with a total size of 20 blocks, according to Rule (12). The algorithm *Tree Initialization* then classifies $a[2][3][2]$ as two initial trees, indexed by $a[0]$ and $a[1]$. The elements in both groups are then recursively initialized by $init_{arrray}$ in sequence. For example, in the recursion of $a[0]$, $id_{array}$ calculates the size of the group indexed by $a[0]$ as $\left(size_1' + size_1' * (size_2' + size_2' * size_3')\right) / \; size_1' = 10$ blocks from $Block_0$ to $Block_9$. Step (2) allocates the memory blocks. Because $a[0]$ is also the beginning address of the whole array, it is allocated to $Block_0$ (deep blue block in Step (2) of Figure 7). To allocate the memory blocks of the second dimension, $init_{arrray}$ must proceed to the next level and recursively initialize the sub-groups indexed by $a[0][0]$, $a[0][1]$, and $a[0][2]$. In Step (3), the information of the group indexed by $a[0][0]$ is stored in $Block_1$, which requires $(size_2' + size_2' * size_3') \; / \; size_2' = 3$ blocks from $Block_1$ to $Block_3$ (green block in Step (3) of Figure 7). To allocate the memory blocks of the third dimension, $init_{arrray}$ continues the deep recursion in Steps (4) and (5), which initialize $a[0][0][X], X \in \{0,1\}$. The elements in $a[0][0][X]$ are of type $\tau_{final}$, so this is the leaf-node level with a group size of $size_3' \; / \; size_3' = 1$. In other words, $a[0][0][0]$ and $a[0][0][1]$ are single-element groups. Both groups are contiguously stored in $Block_2$ and $Block_3$ (orange block in Step (4) of Figure 7). Steps (6)–(8) restore the memory state $m_{state_0}$ to the recursion level of the second dimension, and repeat Steps (3)–(5) for $a[0][1][X]$. This process repeats for the remaining groups, until the whole array information has been stored into respective memory blocks. The final structure of $a[2][3][2]$ in the memory space, from $Block_0$ to $Block_{19}$, is shown in Figure 8. Here, the left column (9) is the real structure and the right column (10) is the group classification.

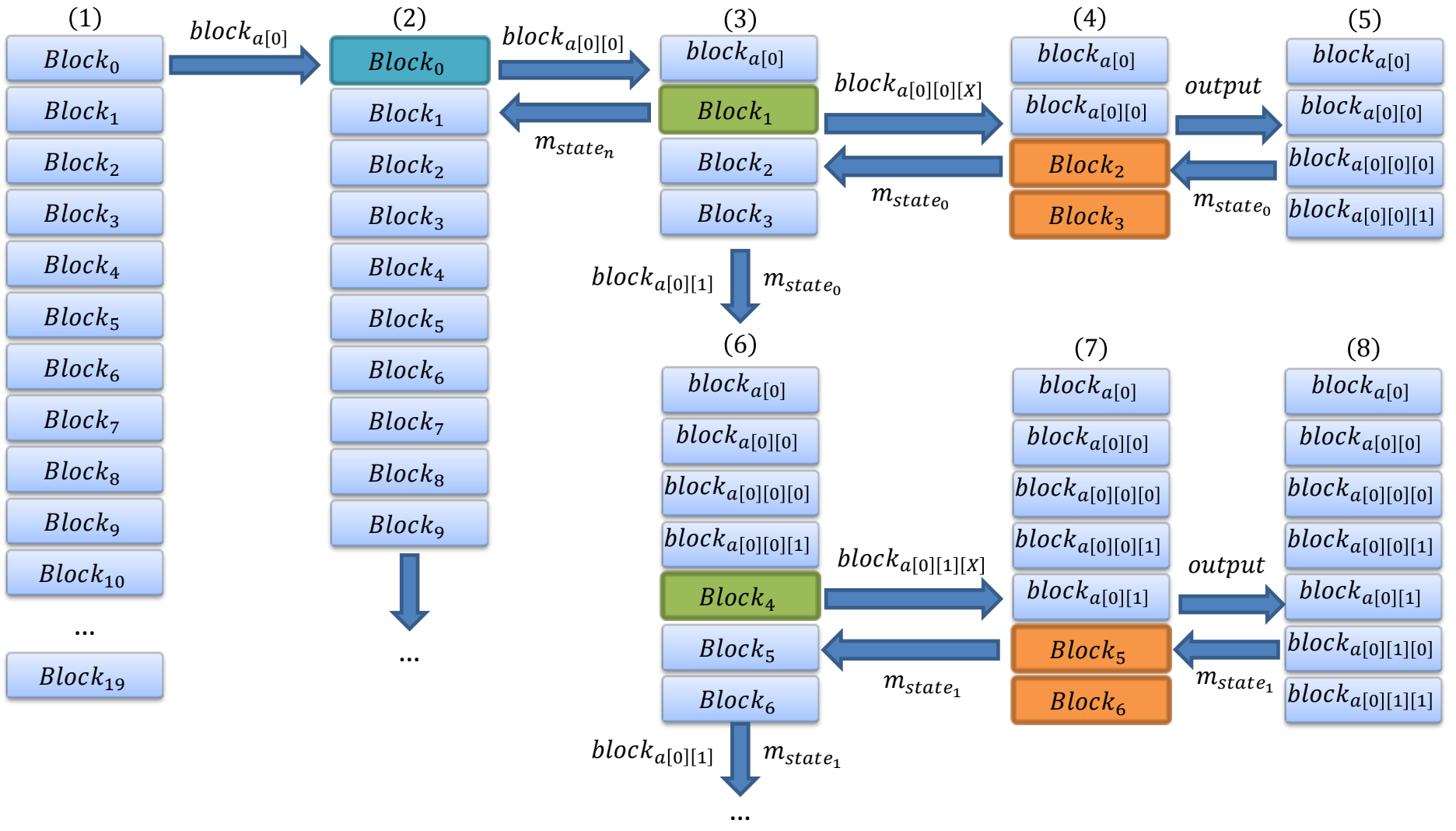

Figure 7. Array value initializing process (see text for details)

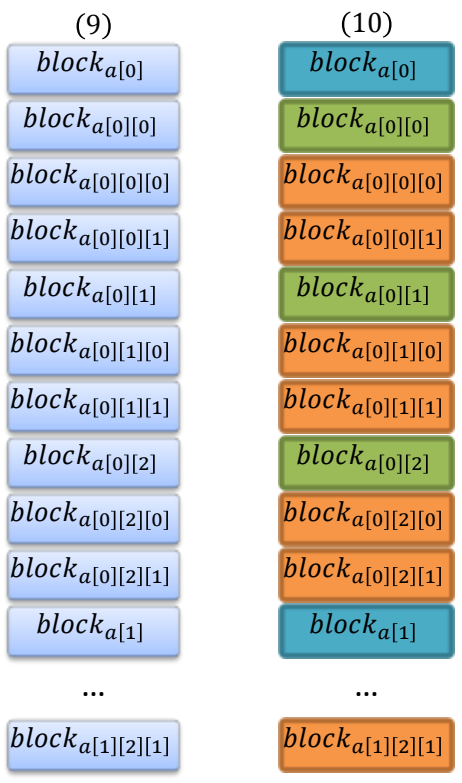

Figure 8. Final structure of the $a[2][3][2]$ array in GERM memory space

The $id_{search}$ can be implemented by a similar algorithm. However, as $id_{search}$ directly locates the indexed group rather than searching each group, its core procedure is $address_{offset}(+, offset, init_{addr})$. The offset formula is given as Rule 15. For example, when locating the block of $a[0][1][1]$, the offset is calculated as $0 * 10 + 1 * 3 + 1 * 1 + 2 = 6$, and the initial address is $Block_0$. Therefore, the information of $a[0][1][1]$ is stored in $Block_6$.

$$offset \equiv \sum_{i=1}^{n} index_i * group_{size_i} + (n-1) \ . \tag{15}$$

Generally, if $valid_{array}(\tau) \wedge alloc(m_{state}, array_{size}) = Some\ init_{addr}$, the array space can be initialized by Algorithm 1, called *Tree Initialization*.

Algorithm 1. Algorithm of the $init_{array}$ function

Algorithm *Tree Initialization*

**Function:** *Fixpoint* $init_{array}$

**Input:** Initial *K* steps, initial memory state $m_{state}$, current initial address, array type, current environment *env*;

**Output:** The final memory state signed with optional type;

$\boldsymbol{Step_0}$: if $\tau = Tarray\ [id\ \tau_{recursive}] \wedge index_i < group_{size_i}$, then move to $\boldsymbol{Step_1}$, else return to $Step_2$;

$\boldsymbol{Step_1}$: $\boldsymbol{let}\ m'_{state} \coloneqq write_{dir}(m_{state}, name, v_{array_i})$,

$init_{array}\left(K-1, m'_{state}, \left(address_{offset}(+, offset, name)\right), \tau_{recursive}, env\right)$;

$\boldsymbol{Step_2}$: if $\tau \in \tau_{final} \wedge index_i < group_{size_i}$, then move to $\boldsymbol{Step_3}$, else return $m_{state}$;

$\boldsymbol{Step_3}$: $\boldsymbol{let}\ m''_{state} \coloneqq write_{dir}(m_{state}, name, v_{array_i})$,

$init_{array}\left(K-1, m''_{state}, \left(address_{offset}(+, offset, name)\right), \tau_{recursive}\right)$;

After running this algorithm, the symbolic execution in FEther more accurately simulates the initialization and allocation behavior of an n-dimensional array in real hardware than other formalizations using the *list* datatype. An array can be abstracted by a number of interesting algorithms, such as tree structure mapping [26] or graphic mapping [27], but the advantages of these algorithms are partially offset by disadvantages. For example, although they can represent an infinite memory space, their specifications and formal structures are very complex and difficult to extend. Moreover, to modify an array element, an operation must search each node one by one, and the overflow problem is difficult to check without a dependent type. In an algorithm based on the GERM memory model, the array is stored in a fixed-size contiguous memory space without assistance by a dependent type [28]. Verifiers can formally simulate the address offsetting process, check the array overflow problem by checking the head Flag stored in the memory block, and modify an array block directly by indexing the respective memory address. Consequently, the verification process becomes easier and more accurate.

Assuming that the current logic context based on GERM has sufficient logic memory space, and that each identifier has a valid and free address, $init_{var}$ represents the first time of setting the indexed memory block, and $write_{dir}$ is always successful. The variable declaration semantics in this scenario are defined in Table A.16 of A.

The semantics of the structure datatype declaration are defined in Table A.17 of A. By Convention 1, the structure declaration at the statement level declares a new structure type with address identifier $str_\tau$. The field member list of $str_\tau$ is $mems_\tau$. As an example, Figure 9 defines the built-in *address* datatype of Solidity rewritten by Lolisa. The *_0xaddress* is the $str_\tau$, and the remaining fields are the $mems_\tau$. The $\mathcal{ESS}_{str}$ records the struct type information directly into the memory block with address $str_\tau$.

*Address* $\stackrel{\text{def}}{=}$ $Str_{type}$ *_0xaddress* ($str_{mem}$ *TInt* (*Nvar addr*)
$str_{mem}$ *TInt* (*Nvar balance*)
$str_{mem}$ (*Tfid* (*Some _Send*)) (*Nvar send*)
$str_{mem}$ *TInt* (*Nvar gas*) ($str_{nil}$))))) *_0xaddress 2 occupy*)

Figure 9. Address type declaration in Solidity, and its equivalent special struct type in Lolisa syntax

In Lolisa, a function call statement unfolds the function body stored in the respective memory address. The semantics of a function call $\mathcal{ESS}_{call}$ are given in Table A.18 of A. In the first step, the function call attempts to read the function declaration statements stored in the respective memory address. If the readout is successful, the second step sets the current execution environment level to 0, and (with the assistance of $set_{env}$) sets the domain as the called function identifier. In the final step, the function body is executed with the new *env'*.

The semantics of the *for* and *while* loops in Lolisa are also similar to those of other well-known languages. The semantics incorporate four conditions: 1) continuation to the next loop, 2) exiting from the current loop during a loop failure, 3) breaking from the current loop, and 4) exiting from the current loop after a *None* message. Because FEther incorporates the BMC model, the loops are unfolded as special sequence statements. Specifically, $\mathcal{ESS}_{for}$ and $\mathcal{ESS}_{while}$ are simplified as two-step functions. First, they judge condition $e$ in its current memory state. Second, if the result of $e$ is true, they evaluate the loop body and transmit the new memory state and loop body into $\mathcal{ESS}$; otherwise, they transmit the current memory state and the next statement into $\mathcal{ESS}$. The subsidiary function $get_{bool}$ is an intermediate interface that avoids the IRE problem. The pattern matching mechanism of Coq requires the listing of all *value*-type constructors, even though the GADTs of Lolisa guarantee that if $\mathcal{ESE}_{rexpr}$ is evaluated successfully, the return value of $e$ takes the form $Some\ Bool(b, env, b_{infor})$. To hide the useless information, $get_{bool}$ extracts the native Boolean value $b :: bool$. In this way, the branches in the current context need only to list the *true* and *false* constructors. The semantics of the *for* and *while* loops are defined in Tables A.19 and A.20 of A, respectively.

Modifier declarations are special function declarations requiring three steps, and including a single limitation. The parameter values are set by the $set_{par}$ predicate. As defined by Table A.21 in A, the first step initializes and sets the parameters. The second step stores the modifier body into the respective memory block, and the third step attempts to initialize the return address $\Lambda_{fun}$. Under the rules of Solidity, the modifier body can return the checking flag, but cannot change the memory states. Therefore, in FEther, we add a special Boolean-type memory block in the GERM framework, indexed by *_0xmodifer*. If the modifier checking is successful, the block is set to *true* and assigned as $\sigma_{mtrue}$; otherwise, it is set to *false* and assigned as $\sigma_{mfalse}$, meaning that other blocks cannot be modified.

To guarantee the type safety, Lolisa separately defines the single- and multi-return value functions. As shown in Table A.22 of A, however, we combine them such that the return type and modifier limitation are both defined as lists. The evaluation is completed by the $repeat$ function. Unlike modifier semantics, the function semantics check the modifier limitations restricting the function. Specifically, all modifiers restricting the function are executed before the function is invoked. If the $modif_{chck}$ result of a modifier evaluation is true, the function is executed; otherwise, it is terminated. Particularly, if $modif_{chck}$ finds a modified memory state, the execution is discarded.

The conditional-statement semantics (see Table A.23 of A) are similar to those of other well-known languages. Under the following rules, $\mathcal{ESE}_{rexpr}$ returns a Boolean result.

Assignment-statement semantics are based on the expression-evaluation semantics. If the result of evaluating an r-value expression is a function pointer generated by a field access, then the return values are evaluated by function call semantics. The semantics of assignment statements are defined in Table A.24 of A.

The returned expressions are evaluated by expression semantics. Valid results are written into their respective $\Lambda_{fun}$. The $\mathcal{ESS}$ then converts the domain of the current environment into a super-domain. Obviously, any statements after the return statement will be stopped by $env_{check}$. Note that because of the GADT, each expression has a different datatype that cannot be stored in a normal list. Therefore, the $es$ is defined as a heterogeneous list storing expressions of different types, and $eval_{exprs}$ is a subsidiary function that evaluates all expressions by repeating the $\mathcal{ESE}_{rexpr}$ process. The semantics of return and multiple-return statements are defined in Tables 25 and 26 of A, respectively.

Finally, Table A.27 of A defines the skip statement *Snil* (which prohibits changes in the skipped part), the throw statement *Throw* (a special kind of Solidity statement that stops the current program and sets the memory state as $\sigma_{init}$), and the function stop statement *FunStop* (a conventional Lolisa statement that represents the completed execution of all statements in the function bodies with no return statements, and resets the current environment).

### 4.3 The FEther Parser

To analyze the syntactic units of *FRWprograms*, the semantics must be integrated into a parser that is easily implemented on the ISA. As shown in Figure 2, the parser has three layers for parsing the three syntax layers. The functions of these layers are validating the environment, deconstructing the input syntactic units, mapping the syntactic units $\mathcal{S}_i$ into the respective semantics $\mathcal{ES}_i$, and transmitting the information stored in the $\mathcal{S}_i$ to the $\mathcal{ES}_i$. As an example, consider the value layer in Table 6. First, the $\mathcal{ESV}$ checks the *K* limitation. It then deconstructs the input value $v$ into specific constructors by pattern matching. Finally, the logic data are transmitted into their respective semantics.

Therefore, the parsers can be summarized as the typing judgements 16 and 17, where $valid$ denotes the validation process.

$$\frac{\mathcal{E} \vdash env,fenv \quad M \vdash \sigma,b_{infor} \quad valid(K,env,fenv)=true \quad \mathcal{S}_i \approx \mathcal{ES}_i}{\mathcal{E},M,\mathcal{F} \vdash \mathcal{S}_i(args) \Rightarrow \mathcal{ES}_i(\sigma,env,fenv,b_{infor},args)} \quad (16) \qquad \frac{\mathcal{E} \vdash env,fenv \quad M \vdash \sigma,b_{infor} \quad \mathcal{F} \vdash K \quad valid(K,env,fenv)=false \quad \mathcal{S}_i \approx \mathcal{ES}_i}{\mathcal{E},M,\mathcal{F} \vdash \mathcal{S}_i(args) \Rightarrow None} \quad (17)$$

| |
|---|
| $\mathcal{ESV} :: \mathbb{Z} \to (\forall\, \tau{:}\, type, val\ \tau) \to memory \to Blc \to Env \to option\ value$<br>$\mathcal{ESV} \equiv$<br>$\lambda\,(K{:}\,\mathbb{Z}).\lambda\big(v{:}\,(\forall\,\tau{:}\,type, val\ \tau)\big).\lambda\,(env{:}\,Env).\lambda\,(\sigma{:}\,memory).\lambda\,\big(b_{infor}{:}\,Blc\big).$<br>$\{\mid valid \longmapsto$<br>$\{\mid Vconst(\tau,n) \longmapsto \Big(\lambda\,\big(b_{infor}{:}\,Blc\big).\mathcal{ESV}_{const}\big(n,env,b_{infor}\big)\Big)$<br>$Varray(index,\tau,name) \longmapsto \Big(\lambda\,\big(b_{infor}{:}\,Blc\big).\mathcal{ESV}_{array}\big(env,\sigma,b_{infor}\big)\Big)$<br>...<br>$\mid\}.v$<br>$invalid \longmapsto None$<br>$\mid\}.K_{check}(K)$ |

Table 6. Simple example of the value-layer parser

The information in $\mathcal{S}_i$ needs to be partially preprocessed before transmission to $\mathcal{ES}_i$. First, we must check whether the constructor of $\mathcal{ES}_{rexpr_{eaddr}}$ is $Efun$. If true, we must transmit the respective $\Lambda_{fun}$ instead of $name$. This action is recorded as $\{\mid\ Efun(oaddr,\tau,*) \ \longmapsto\ read_{chck}\big(\sigma,env,b_{infor},\Lambda_{fun}\big) \mid\}.e$. As mentioned above, after evaluating $\mathcal{ESS}_{re}$ and $\mathcal{ESS}_{res}$, we must then change the current environment into a super-environment for stopping the function execution. Moreover, as semantics such as $\mathcal{ESS}_{call}$, $\mathcal{ES}_{rexpr_{bop}}$ and $\mathcal{ES}_{rexpr_{uop}}$ recursively invoke $\mathcal{ESE}_{rexpr}$, the specific $\mathcal{ESE}_{rexpr}$ and $\mathcal{ESS}$ must be defined as recursive functions. Finally, the parser statement level integrates the two lower levels, and also defines the entering point of FEther (see the *FEther_enter_point* algorithm given as Algorithm 2).

Algorithm 2. Algorithm of the FEther entering point

| |
|---|
| Algorithm FEther_enter_point<br>**Function:** *Fixpoint* FEther<br>**Input:** Initial *K* steps, optional initial memory state $[\![m_{state}]\!]$, current environment *env*, and super-environment *fenv*; initial arguments $args$, and valid $FRWprogram$;<br>**Output:** The final memory state signed with optional type;<br>$\boldsymbol{Step_0}$: if $env_{check}(env,fenv) = true \wedge [\![m_{state}]\!] \neq None$, then move to $\boldsymbol{Step_1}$, else return $[\![m_{state}]\!]$;<br>$\boldsymbol{Step_1}$: if $FRWprogram = s_0 :: s_1$, set $env' = set_{env}(s_0 :: s_1, env)$ and move to $\boldsymbol{Step_2}$, else return $[\![m_{state}]\!]$;<br>$\boldsymbol{Step_2}$: if $\mathcal{S}_0 \approx \mathcal{ES}_0$ then $(K,m,env,fenv,\mathcal{ES}_0) \Downarrow_{P(s_0)} \xrightarrow{yields} [\![m'_{state}]\!]$ and move to $\boldsymbol{Step_3}$, else return $None$.<br>$\boldsymbol{Step_3}$: $FEther([\![m'_{state}]\!], env', fenv, args, s_1)$ |

The rules governing the execution of a Lolisa program in FEther are defined by the rules EXE-F and EXE-IF below, where the symbol $\infty$ refers to infinite execution, and *T* is the termination condition set of a finite execution.

$$\frac{\mathcal{E} \vdash env,fenv \quad M \vdash \sigma,b_{infor} \quad \mathcal{F} \vdash opars \quad \mathcal{E},M,\mathcal{F} \vdash P(stt) \quad \mathcal{E},M,\mathcal{F} \vdash lib \quad env=set_{gas}\big(init_{env}(P(stt))\big) \quad fenv=init_{env}(P(stt)) \quad \sigma=init_{mem}(P(stt),lib)}{\mathcal{E},M,\mathcal{F} \vdash FEther\big([\![m'_{state}]\!],env',fenv,args,P(stt)\big) \xrightarrow{execute,T} \langle \sigma',env,fenv \rangle} \ \text{(EXE-F)},$$

$$\frac{\begin{array}{c}\mathcal{E}\vdash env,fenv \quad M\vdash\sigma,b_{infor} \quad \mathcal{F}\vdash opars \quad \mathcal{E},M,\mathcal{F}\vdash P(stt) \quad \mathcal{E},M,\mathcal{F}\vdash lib \\ env=set_{gas}\big(init_{env}(P(stt))\big) \quad fenv=init_{env}\big(P(stt)\big) \\ \sigma=init_{mem}(P(stt),lib)\end{array}}{\mathcal{E},M,\mathcal{F}\vdash FEther\big([\![m'_{state}]\!],env',fenv,args,P(stt)\big)\xRightarrow{execute,\infty}\langle\sigma',env',fenv\rangle\vee env'.(gas)\to\big(\neg fenv.(gasLimit)\big)\xRightarrow{execute,T}\langle\sigma',env',fenv\rangle}\ \text{(EXE-IF)}.$$

### 4.4 Automation tactics

Automated theorem proving is a core topic in formal verification research. Many higher-order theorem-proving assistants provide tactics or similar mechanisms that simplify the program evaluation process and construct proofs automatically. With manual modeling technology, different formal models with significantly different structures and verification processes can be constructed in various programs. Hence, designing a set of tactics that automatically verifies models in different programs is nearly impossible.

The above problem is circumvented by FEther. According to EVI theory, the FEther symbolic execution corresponds to both the function evaluation and the program verification (see Rule 18). In other words, it unifies the verification processes of different programs in higher-order theorem-proving assistants by simplifying the program evaluation process in FEther. Because the situations of FEther execution constitute a fixed and finite set $\{s_0, s_1, \dots, s_m\}$, we can design sufficiently many sub-tactics for all situations. Exploiting this advantage and assisted by the *Ltac* mechanism, we designed primitive automatic tactics for the FEther. The tactic strategy model is constructed from three parts: memory operating, *K* costing and semantics simplifying.

$$\Omega, M, \mathcal{F} \vdash_{ins} P_{exe} \equiv P_{eval} \equiv P_{verify}\,. \tag{18}$$

The workflow is defined in Figure 10. When the proof universe of Coq is open, the $observe$ function scans the current context $C$ to obtain the current goal. In sequence, each part attempts to capture the operation characteristic of the current goal and choose the matching tactics. The selected tactics are combined into a solution tactic $Ltac_i$ that solves the goal in TCOC. The new context $C'$ is compared with $C$ in $context_{dec}$. If $C'$ and $C$ are identical, the current tactics cannot solve the goal automatically, and the tactic model is terminated. Otherwise, the tactic model continuously attempts to simplify the goal of $C'$.

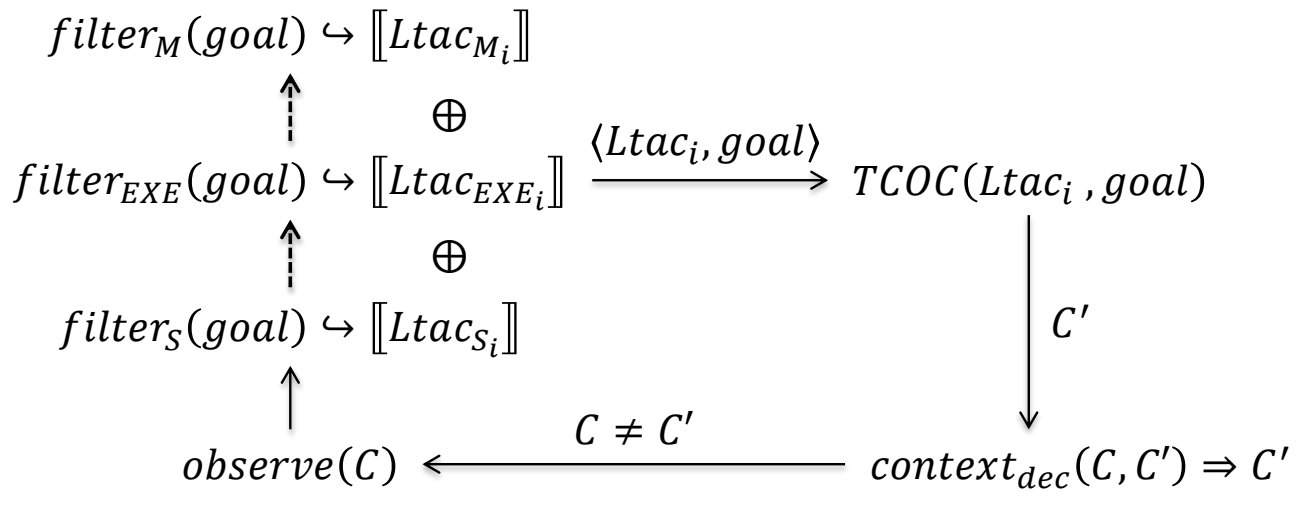

Figure 10. An abstract automatic tactic working process

Algorithm 3 states the *unfold_modify* tactic, a sub-tactic of the memory operation part. This sub-tactic captures parts of the operation characteristic of the $write_{dir}$ function, and evaluates the scanned $write_{dir}$ using basic built-in tactics.

Algorithm 3. Sub-tactic for capturing the operation characteristic of the $write_{dir}$ function

```
Ltac unfold_modify:=
  match goal with
    | [ |- context [?Y (?X: memory) (?Z: value)]] ⇒ unfold Y in *; cbn in *
  end.
```

The average ratios of contract size to proof size are shown in Table 7. Smart contracts exceeding 500 lines were excluded from this analysis, because the size of large contracts were limited by the *gas* cost. The second and third columns of Table 7 list the ratios using Coq's built-in tactics and our automatic tactics, respectively. Obviously, the automatic tactics reduce much of the proof workload. Moreover, according to our experimental results, the ratio floats in a range is influenced by the complexity of the target contract. Specifically, the ratios obtained by the built-in tactics range from approximately $-0.5$ to $+10.0$, whereas those of the automatic tactics range from approximately $-0.1$ to $+0.3$. Therefore, the automatic tactics possess a better universal property than the directly applied built-in tactics.

| Contract size | Built-in tactics | Automatic tactic |
|---|---|---|
| **$\leq$ 100 lines** | 1.27 | 0.16 |
| **$\leq$ 200 lines** | 2.33 | 0.13 |
| **$\leq$ 300 lines** | 5.26 | 0.17 |
| **$\leq$ 400 lines** | 7.5 | 0.22 |
| **$\leq$ 500 lines** | 10.3 | 0.17 |

Table 7. Ratios of proof size to contract size in theorem-proving tactics

### 4.5 Self-correctness Certification

The FEther interpreter is entirely constructed in Coq, which confers a natural advantage over other program verifications and analysis tools. The core of Coq is the trusted computation base (TCB) [24], which satisfies the de Bruijn criterion. In almost all program analysis tools, TCB self-verification is arguable and paradoxical, so whether the TCB of a program verification (analysis) tool satisfies the de Bruijn criterion is an important indicator of the trustworthiness of the verification.

The correctness of FEther is certified by its consistency between the relational and computational definitions, the correctness of its essential properties, and the meta-properties of its semantics.

First, we must prove that the operational semantics [30] of Lolisa (the inductive relational forms) are equivalent to the operational semantics (the executable function forms). As desired in the CompCert project [26], we check whether each evaluation in the relation semantics corresponds to the symbolic execution in the executable semantics. For this purpose, we construct a simulation diagram. Under identical conditions, the relational and executable semantics must have the same observable effect (same traces of the evaluation process). This requirement is embodied in the following *simulation diagram* theory.

**Theorem** (simulation diagram) Let $\mathcal{E}, M, \mathcal{F} \vdash_{ins} \sigma, opars, env, fenv, b_{infor}$ be the initial evaluation environment, and let $R_{eq}$ represent the equivalence relationship between two terms. Then, any relational semantic $S_{rel}$ and executable semantic $S_{exe}$ must satisfy the following simulation diagram:

$$
\begin{array}{ccc}
S_{rel} & \xrightarrow[\langle \sigma, env, fenv, b_{infor}, opars, \Downarrow_{S_{rel}} \rangle]{\mathcal{E}, M, \mathcal{F} \vdash_{ins} \sigma, opars, env, fenv, b_{infor}} & [\![ result ]\!] \\
\updownarrow R_{eq} & & \updownarrow R_{eq} \\
S_{exe} & \xrightarrow[S_{exe}(n, \sigma, opars, env, fenv, b_{infor})]{\mathcal{E}, M, \mathcal{F} \vdash_{ins} \sigma, opars, env, fenv, b_{infor}} & [\![ result' ]\!]
\end{array}
$$

Second, we must certify the correctness of the foundation behavior of the executable semantics. As a simple instance, we construct Lemma *test_lemma_if_false*, which certifies the correctness of the following execution: For all statements $s$ and $s'$, if the *if* statement condition is false, FEther must execute the statement $s'$ of the false branch. By a similar process, we certified that almost all of the executable semantics exhibit standard behaviors.

**Lemma** (test_lemma_if_false)

$$\forall\ if_{false}\ if_{state}\ s\ s'\ n\ env\ pass,$$
$$if_{false} = Econst\ (Vbool\ false) \rightarrow$$
$$if_{state} = (If\ if_{false}\ s\ s') \rightarrow$$
$$n > 0 \rightarrow$$
$$(FEther\ n\ init_m\ pass\ env\ env\ if_{false}) = (FEther\ n\ init_m\ pass\ set_{gas}(env)\ env\ s').$$

Finally, we prove the meta-properties of these semantics. The most basic properties in each layer are the *progress* and *preservation* properties, which maintain the static-type safety of the specification. For example, the *progress* and *preservation* of the expression layer are defined in Lemma *expression type safety*. Because Lolisa is a strongly typed language defined in terms of GADTs, the *progress* and *preservation* properties of expressions are easily proven by simplifying the semantics function. The *progress* and *preservation* properties of other layers are certified similarly. Besides the meta-properties, we proved the execution determinism of all semantics in Coq. The Lemma *execution determinism* is one example of the relevant proofs.

**Lemma** (expression type safety)

1. If $e\colon expr_{\tau_0\,\tau_1}$ and $e \longmapsto e'$, then $e'\colon expr_{\tau_0'\,\tau_1}$.
2. If $e\colon expr_{\tau_0\,\tau_1}$, then either $e(v)$ or some $e'$ exists such that $e \longmapsto e'$.

**Lemma** (execution determinism)

$$\forall\, s\; m\; [\![m_{final}]\!]\; [\![m'_{final}]\!]\; n\; env\; pass,$$
$$FEther\; n\; m\; pass\; env\; env\; s = [\![m_{final}]\!] \rightarrow$$
$$FEther\; n\; m\; pass\; env\; env\; s = [\![m'_{final}]\!] \rightarrow$$
$$m_{final} = m'_{final}$$

At present, the core functions have been completely verified. The correctness certification includes 74 theorems and lemmas, and approximately 4000 lines of Coq proof code.

## 5 Formal verification of smart contract by FEther

To demonstrate the power of FEther in real-world practice, this section verifies the smart contract in a case study that demonstrates our proof engine and its benefits. The experimental environment was five identical personal computers with equivalent hardware of 8 GB memory and a 3.20 GHz CPU. All computers were run on Windows 10 and CoqIDE 8.8. For readers' benefit, the code of the examples is given in Appendix B.

### 5.1 Case Study: Hybrid Verification

As a simple example, we consider the *wallet* function encoded in Appendix B. This function, which executes initial coin offering, is a segment of the Solidity contract extracted from the contract demonstration [2].

First, the smart contract in its formal version was translated line-byline into Lolisa. The result is shown in Figure 11. Among the most important functions of *wallet* is the application time validation. Clearly, the contract will be discarded if the current time *now* are below *privilegeOpen* or above *privilegeClose*.

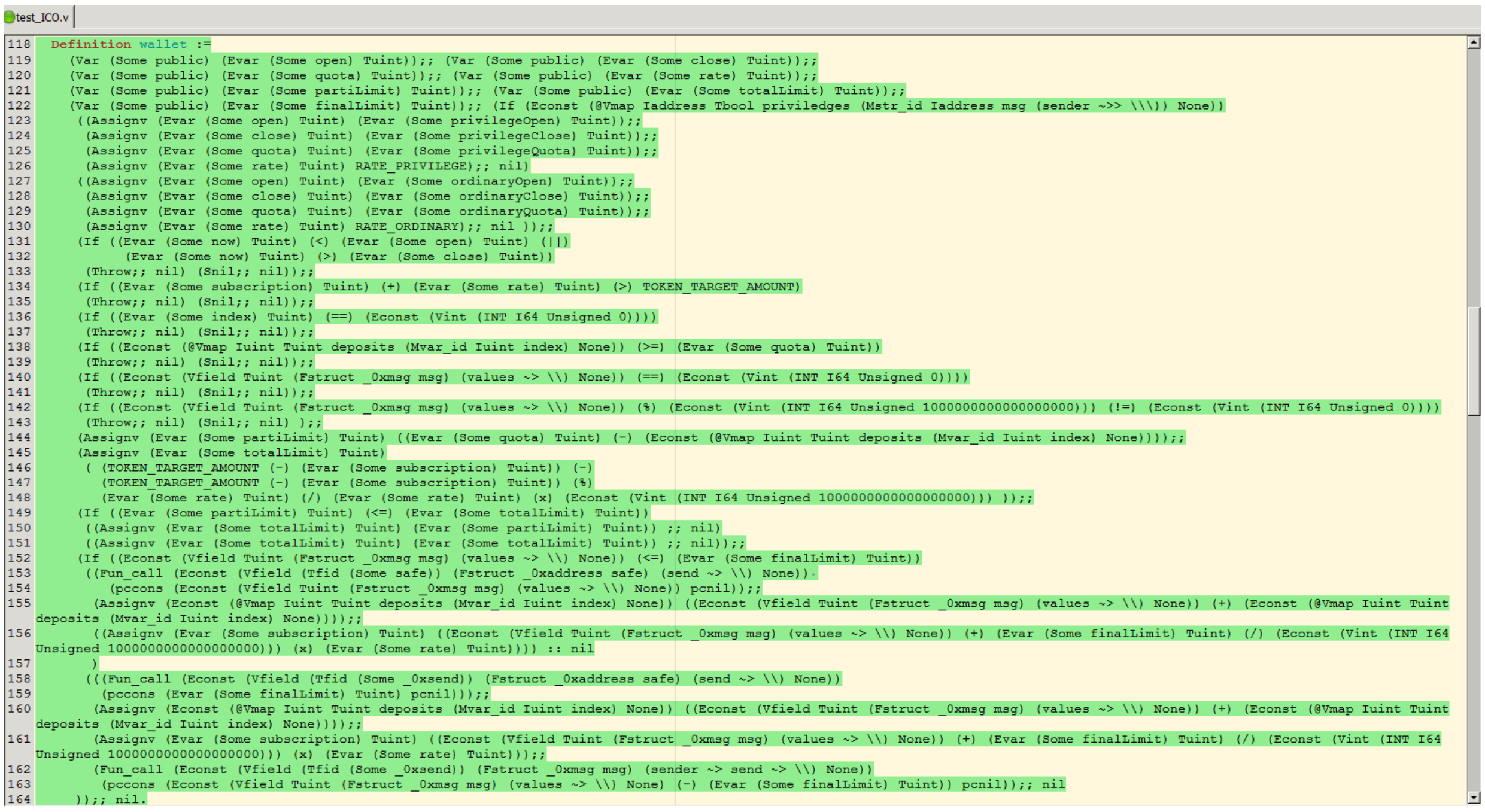

```
Definition wallet :=
   (Var (Some public) (Evar (Some open) Tuint));; (Var (Some public) (Evar (Some close) Tuint));;
   (Var (Some public) (Evar (Some quota) Tuint));; (Var (Some public) (Evar (Some rate) Tuint));;
   (Var (Some public) (Evar (Some partiLimit) Tuint));; (Var (Some public) (Evar (Some totalLimit) Tuint));;
   (Var (Some public) (Evar (Some finalLimit) Tuint));; (If (Econst (@Vmap Iaddress Tbool priviledges (Mstr_id Iaddress msg (sender ~>> \\\)) None))
    ((Assignv (Evar (Some open) Tuint) (Evar (Some privilegeOpen) Tuint));;
     (Assignv (Evar (Some close) Tuint) (Evar (Some privilegeClose) Tuint));;
     (Assignv (Evar (Some quota) Tuint) (Evar (Some privilegeQuota) Tuint));;
     (Assignv (Evar (Some rate) Tuint) RATE_PRIVILEGE);; nil)
    ((Assignv (Evar (Some open) Tuint) (Evar (Some ordinaryOpen) Tuint));;
     (Assignv (Evar (Some close) Tuint) (Evar (Some ordinaryClose) Tuint));;
     (Assignv (Evar (Some quota) Tuint) (Evar (Some ordinaryQuota) Tuint));;
     (Assignv (Evar (Some rate) Tuint) RATE_ORDINARY);; nil ));;
    (If ((Evar (Some now) Tuint) (<) (Evar (Some open) Tuint) (||)
          (Evar (Some now) Tuint) (>) (Evar (Some close) Tuint))
     (Throw;; nil) (Snil;; nil));;
    (If ((Evar (Some subscription) Tuint) (+) (Evar (Some rate) Tuint) (>) TOKEN_TARGET_AMOUNT)
     (Throw;; nil) (Snil;; nil));;
    (If ((Evar (Some index) Tuint) (==) (Econst (Vint (INT I64 Unsigned 0))))
     (Throw;; nil) (Snil;; nil));;
    (If ((Econst (@Vmap Iuint Tuint deposits (Mvar_id Iuint index) None)) (>=) (Evar (Some quota) Tuint))
     (Throw;; nil) (Snil;; nil));;
    (If ((Econst (Vfield Tuint (Fstruct _0xmsg msg) (values ~> \\) None)) (==) (Econst (Vint (INT I64 Unsigned 0))))
     (Throw;; nil) (Snil;; nil));;
    (If ((Econst (Vfield Tuint (Fstruct _0xmsg msg) (values ~> \\) None)) (%) (Econst (Vint (INT I64 Unsigned 1000000000000000000))) (!=) (Econst (Vint (INT I64 Unsigned 0))))
     (Throw;; nil) (Snil;; nil) );;
    (Assignv (Evar (Some partiLimit) Tuint) ((Evar (Some quota) Tuint) (-) (Econst (@Vmap Iuint Tuint deposits (Mvar_id Iuint index) None))));;
    (Assignv (Evar (Some totalLimit) Tuint)
     ( (TOKEN_TARGET_AMOUNT (-) (Evar (Some subscription) Tuint)) (-)
       (TOKEN_TARGET_AMOUNT (-) (Evar (Some subscription) Tuint)) (%)
       (Evar (Some rate) Tuint) (/) (Evar (Some rate) Tuint) (x) (Econst (Vint (INT I64 Unsigned 1000000000000000000))) ));;
    (If ((Evar (Some partiLimit) Tuint) (<=) (Evar (Some totalLimit) Tuint))
     ((Assignv (Evar (Some totalLimit) Tuint) (Evar (Some partiLimit) Tuint)) ;; nil)
     ((Assignv (Evar (Some totalLimit) Tuint) (Evar (Some totalLimit) Tuint)) ;; nil));;
    (If ((Econst (Vfield Tuint (Fstruct _0xmsg msg) (values ~> \\) None)) (<=) (Evar (Some finalLimit) Tuint))
     ((Fun_call (Econst (Vfield (Tfid (Some safe)) (Fstruct _0xaddress safe) (send ~> \\) None))
        (pccons (Econst (Vfield Tuint (Fstruct _0xmsg msg) (values ~> \\) None)) pcnil));;
      (Assignv (Econst (@Vmap Iuint Tuint deposits (Mvar_id Iuint index) None)) ((Econst (Vfield Tuint (Fstruct _0xmsg msg) (values ~> \\) None)) (+) (Econst (@Vmap Iuint Tuint
deposits (Mvar_id Iuint index) None))));;
      ((Assignv (Evar (Some subscription) Tuint) ((Econst (Vfield Tuint (Fstruct _0xmsg msg) (values ~> \\) None)) (+) (Evar (Some finalLimit) Tuint) (/) (Econst (Vint (INT I64
Unsigned 1000000000000000000))) (x) (Evar (Some rate) Tuint)))) :: nil
      )
     (((Fun_call (Econst (Vfield (Tfid (Some _0xsend)) (Fstruct _0xaddress safe) (send ~> \\) None))
       (pccons (Evar (Some finalLimit) Tuint) pcnil)));;
      (Assignv (Econst (@Vmap Iuint Tuint deposits (Mvar_id Iuint index) None)) ((Econst (Vfield Tuint (Fstruct _0xmsg msg) (values ~> \\) None)) (+) (Econst (@Vmap Iuint Tuint
deposits (Mvar_id Iuint index) None))));;
      (Assignv (Evar (Some subscription) Tuint) ((Econst (Vfield Tuint (Fstruct _0xmsg msg) (values ~> \\) None)) (+) (Evar (Some finalLimit) Tuint) (/) (Econst (Vint (INT I64
Unsigned 1000000000000000000))) (x) (Evar (Some rate) Tuint)));;
      (Fun_call (Econst (Vfield (Tfid (Some _0xsend)) (Fstruct _0xmsg msg) (sender ~> send ~> \\) None))
       (pccons (Econst (Vfield Tuint (Fstruct _0xmsg msg) (values ~> \\) None) (-) (Evar (Some finalLimit) Tuint)) pcnil));; nil
    ));; nil.
```

Figure 11. Formal version of the *wallet* function

According to EVI theory, verification in the proposed FSPVM is founded on simultaneous Hoare logic and reachability logic. Meanwhile, verification in FEther combines higher-order theorem proving and symbolic execution. By virtue of this hybrid system, programmers can mechanically define the Hoare style properties following the formula abstract (19), where the wildcard "*"represents other specific arguments.

$$\mathrm{P}\{\mathrm{m_{init}}\}\ \mathrm{FEther}(\mathrm{m_{init}}, \mathrm{FRWprograms}, *)\ \mathrm{Q}\{\mathrm{m_{final}}\}. \tag{19}$$

According to the reachability logic, the Hoare logic derivation is equivalent to the trusted operational semantics execution. Therefore, the execution of FEther can be seen as a derivation based on Hoare logic. The inference process is given by expression (20). The specific initial memory state $m_{init}$ is the precondition of the program verification. Guided by the semantics of each statement $\mathrm{c_i}$, FEther logically modifies the current memory state $m_{i-1}$ to a new postcondition $Q_i\{m_i\}$ (i.e., the precondition of $\mathrm{c_i}$). The theorems need only judge wither the final output memory state $m_n$ after executing the final statement matches the correct memory state $m_{final}$. Most importantly, this verification procedure is automated in the proposed FSPVM.

$$\mathrm{P}\{\mathrm{m_{init}}\}\mathrm{c_0} \xrightarrow{\mathrm{FInterpreter(m_{init},c_0)}} \mathrm{Q_0}\{\mathrm{m_0}\}\mathrm{c_1} \xrightarrow{\mathrm{FInterpreter(m_0,c_1)}} \mathrm{Q_1}\{\mathrm{m_1}\}\mathrm{c_2} \twoheadrightarrow \mathrm{c_n Q_n}\{\mathrm{m_n}\} \overset{?}{\leftrightarrow} \mathrm{Q}\{\mathrm{m_{final}}\}. \tag{20}$$

During this process, verifiers can alter the verification patterns (including static, concolic, and selective symbolic execution) by defining the preconditions in different ways. For example, programmers can vary the *wallet* function by the following three approaches.

When the initial arguments are inductively defined with quantifiers such as $\forall$ and $\exists$, the traditional symbolic execution will traverse all cases. If the current time $now$ in the wallet function is outside the range $open$ to $close$, the smart contract must be discarded. The Lemma *no_in_time* defined in Figure 12 underlined in red abstractedly defines $(\mathrm{INT\ I64\ Unsigned}\,?X)$ and $(\mathrm{INT\ I64\ Unsigned}\,?Y)$ by specifying $?X$ and $?Y$ as inductive values representing all possible situations, namely, as $\forall\,(\mathrm{x{:}\,Int})(\mathrm{INT\ I64\ Unsigned\ x})$ and $\forall\,(\mathrm{y{:}\,Int})(\mathrm{INT\ I64\ Unsigned\ y})$, respectively. Using the automatic tactics of FEther, programmers can execute and complete this verification within 3.37 s (see Figure 13).

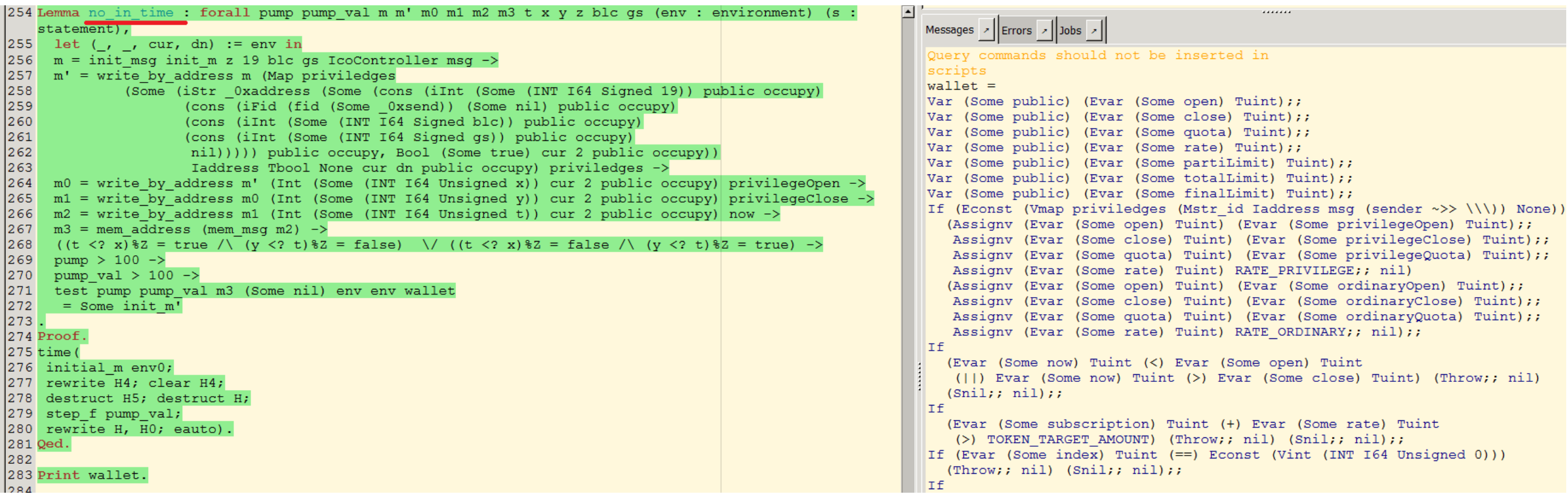

Figure 12. Verification process of *wallet* with abstract symbol arguments

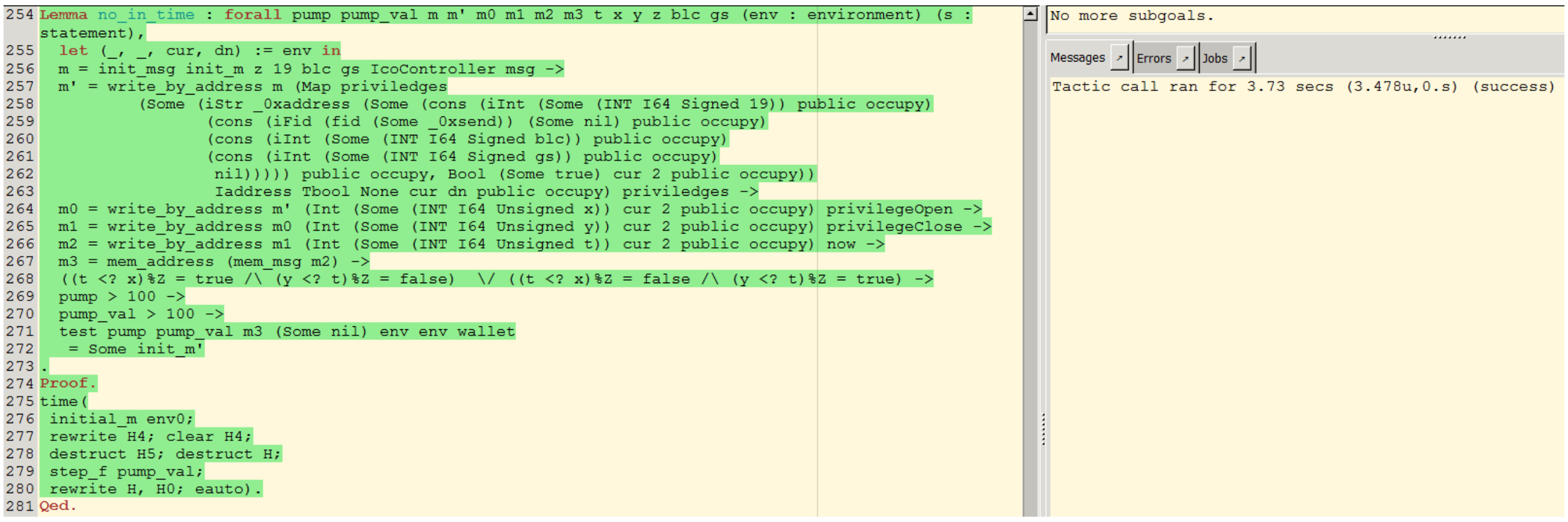

Figure 13. Execution time of verifying *wallet* with abstract symbol arguments

Second, FEther supports concolic symbolic execution that gets real inputs. To accurately simulate execution processes on real world hardware, FEther is built in a virtual execution environment. Therefore, a FEther execution can be regarded as a special dynamic analysis. As shown in Figure 14, the entering points *test* and code *wallet* are unmodified, and *privilegeOpen*, *proviledgeClose* and *now* are replaced with specific values 0, 3, and 4, respectively. The other constraints are still inductively defined as abstract symbols. The function correctness of concolic execution with specific inputs is then proven by the *no_in_time* lemma. Because the inputs are specified, the number of possible execution paths is limited, and the execution time reduces to 1.689 s (see Figure 15). Of course, to test the function of *wallet* within a special test suite, programmers can write an automatic test script that modifies the values of *privilegeOpen*, *provildegeClose* and *now*.

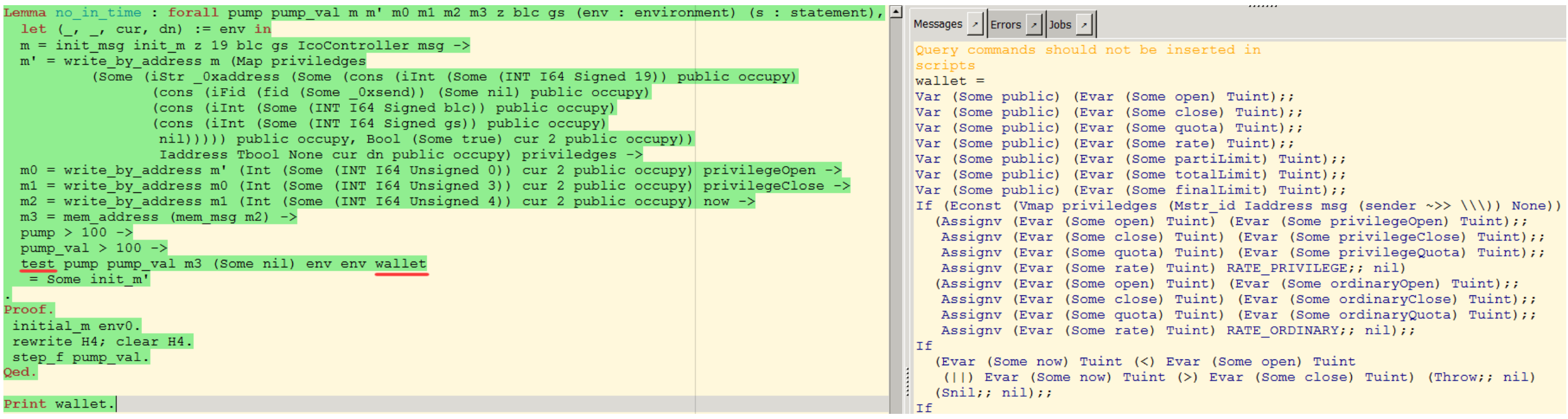

Figure 14. Concolic verification of the *wallet* function

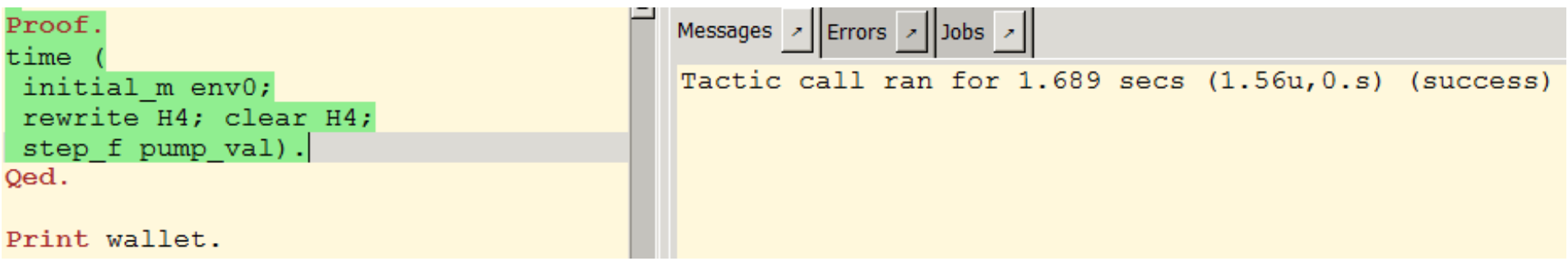

Figure 15. Concolic execution time of *wallet* function

Third, the *wallet* function can be varied by exploiting the selective symbolic execution of FEther. As shown in Figure 16, programmers can extract the core code segment if (now < open || now > close) {throw();}from the *wallet* function and represent it by a new definition such as *msp'* underlined in red, which can be individually verified by the *msp_correct* lemma. After combining the verified *msp'* into the *wallet* function, the verification of the *no_in_time* lemma can be finished by invoking the *msp_correct* lemma. Clearly, the *msp_correct* can also assist the proofs that use the *msp'* code segment.

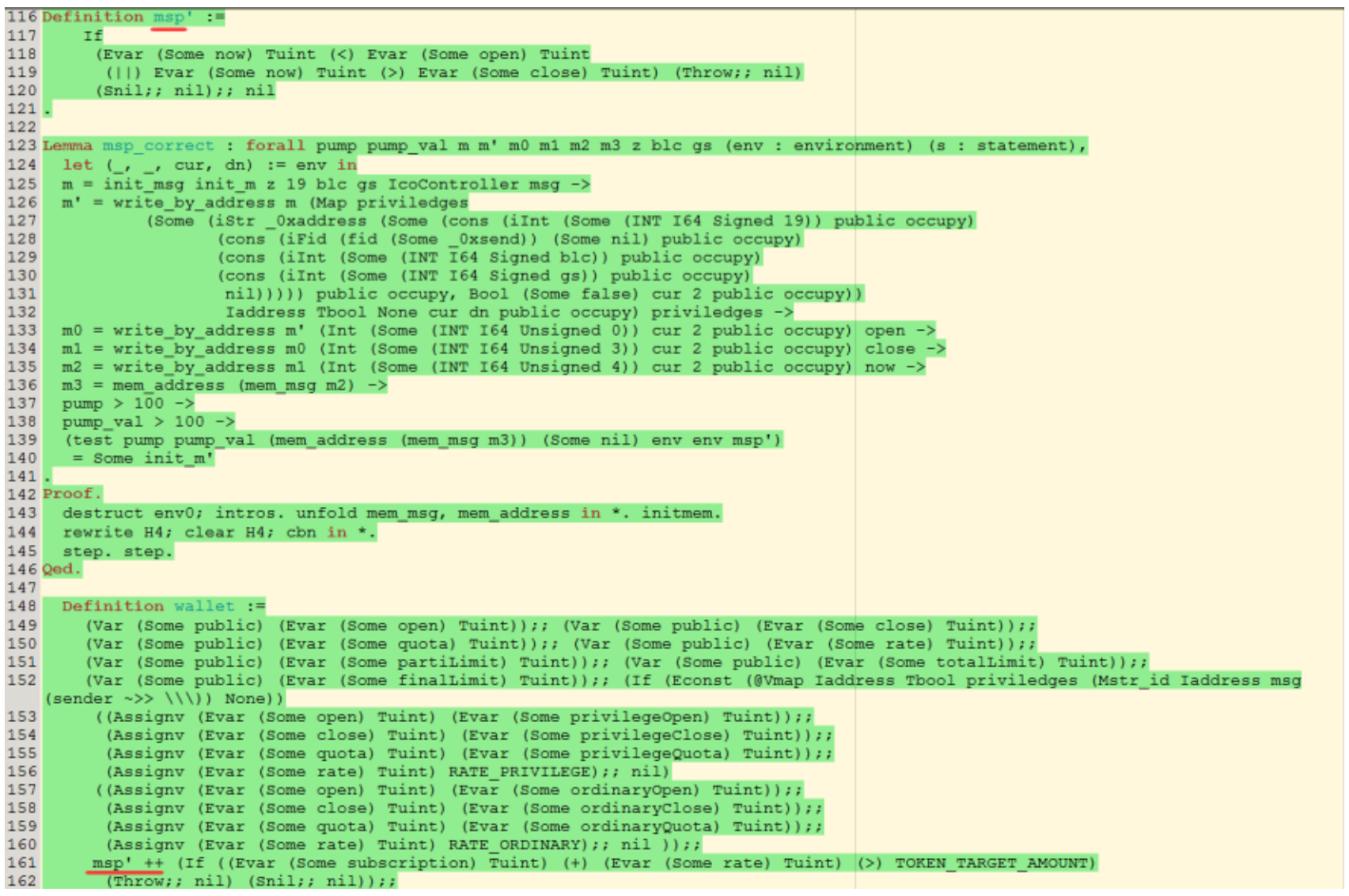

Figure 16. Selective symbolic execution of the *wallet* function

The FEther can also simplify loop proofs. In the standard approach of higher-order theorem proving, program loops are proven by manually identifying the invariants. Searching the loop invariants of simple loops, however, is a tedious process. By combining symbolic execution and higher-order theorem proving, we simultaneously facilitate the use of BMC and the search for loop invariants. Employing BMC, we first limit FEther to *K* or fewer executions of FRWprogram. In general, if *L* executions (where $L \le K$) of an FRWprogram can generate the corresponding final memory state, the loops existing in the FRWprogram can be directly unfolded as a set of identical normal-sequence statements within finite time, as inferred from Rule 21. If the *FRWprogram* fails to generate the corresponding final memory state after *K* executions, we can set the loop statement as a breakpoint (by virtue of the selective symbolic execution) and separate the *FRWprogram* into two parts, denoted as the head and tail parts. Next, we must locate the loop invariants and encapsulate them into an invariant memory state $I\{m_i\}$, which serves as the final memory state of the head part and the initial memory state of the tail part. This procedure is embodied in Rule 21 below.

$$P\{m_{init}\}c_0 \twoheadrightarrow c_i I\{m_i\}(\text{head}) \ \text{and} \ c_i I\{m_i\} \twoheadrightarrow c_n Q\{m_{final}\}(\text{tail}). \quad (21)$$

Under the composition rule of Hoare logic, we have $P\{m_{init}\}c_0 \twoheadrightarrow c_n Q\{m_{final}\}$. In this way, simple loops can be proven automatically, reducing the workload of searching loop invariants. Moreover, complex loops that cannot be verified by model checking and symbolic execution technology can be proved by higher-order theorem-proving technology.

Finally, the FEther provides a debug mechanism for users. Because FEther is developed in the GERM memory model, it provides debug tactics such as *step*, which enables step-by-step debugging of a smart contract. The formal intermediate memory states obtained during the execution and verification process of a Lolisa program using FEther are shown in the proof context (right panel of Figure 17). Programmers can follow the intermediate memory states to locate the bugs.

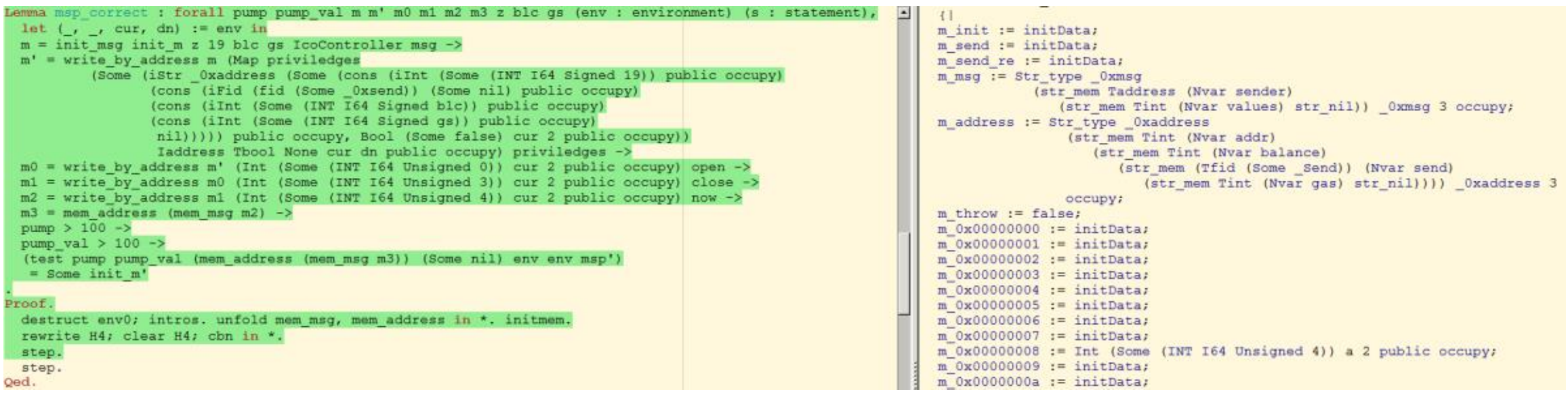

Figure 17. Debugging of the *wallet* function in Coq

Clearly, users of FEther can flexibly choose the most suitable method for verifying their programs.

### 5.2 Feature Comparison Overview

FEther is the first hybrid symbolic execution engine for Ethereum smart contracts. To illustrate the advantages of FEther over the solvers of other tools, we require a compelling benchmark, such as a testing suite or analysis time. Given that FEther is constructed on Coq, however, and directly executes and verifies the Solidity source code of smart contracts rather than compiling Solidity at the bytecode level, such a benchmark is difficult to find. For a fair comparison, we instead compared the presence and absence of various features in FEther and in other tools. The compared features are listed and defined below:

- Spec.: Suitable as a formal specification of the EVM language
- Exec.: Executable on concrete tests
- Certif.: Certifiable self-correctness
- Verif.: Verifiable properties of EVM programs
- Debug.: Provision of an interactive EVM debugger
- Gas.: Tools for analyzing the gas complexity of an EVM program
- Level.: Analysis or verification level of code
- Logic.: Type of essential logic supported
- Hybrid.: Support for hybrid verification methods

| Tool | Spec. | Exec. | Certif. | Verif. | Debug. | Gas. | Level | Logic. | Hybrid. |
|---|---|---|---|---|---|---|---|---|---|
| Yellow Paper | Yes | No | No | No | No | No | None | None | No |
| Lem spec | Yes | Yes | Testing | Yes | No | No | Byte Code | Higher order | No |
| Mythril | Yes | Yes | Testing | Yes | Yes | Yes | Byte Code | First order | No |
| Hsevm | No | Yes | Testing | No | Yes | No | Byte Code | None | No |
| Scilla | Yes | No | Testing | Yes | No | No | Intermediate | Higher order | No |
| Cpp-ethereum | No | Yes | Testing | No | No | No | Byte Code | None | No |
| KEVM | Yes | Yes | Testing | Yes | Yes | Yes | Byte Code | First order | No |
| FEther | Yes | Yes | Verifying | Yes | Yes | Yes | Solidity | Higher order | Yes |

Table 8. Feature comparison of FEther semantics and existing software quality tools

Table 8 overviews the results of the feature comparison. Obviously, only FEther, the core of KEVM, and Mythril support the Spec, Exec, Verif, Debug, and Gas features. The Certif feature of FEther is “verifying” rather than “testing,” which improves the reliability of FEther (at least in theory) over testing methods such as KEVM and Mythril. Moreover, the execution and verification level of FEther is "Solidity" rather than "byte code," which avoids the error risk during compiling. FEther also supports higher-order logic, which improves the expressive ability. Moreover, the fundamental verification theory of FEther is the calculus of inductive construction instead of the satisfiability modulo theories or Boolean satisfiability problem. Therefore, the situations that cannot be evaluated and verified do not exist. Finally, FEther is the only tool that supports hybrid formal verification.

According to our previous experimental results [25], the symbolic execution time of the optimized current version of FEther is approximately 0.03 s per statement when the initial arguments are specified, and approximately 0.07 s when the initial arguments are inductively defined by quantifiers. The execution efficiency of FEther far exceeded that of the interpreters that are developed in Coq in accordance with the standard tutorial developed in Coq. The current version also supports the verification of smart contract models adhering to the Ethereum ERC20 standard.

## 6. Discussion

### A. Contributions

This article overcomes the final challenge noted in our previous work: completing the proof engine of FSPVM-E. We now highlight the significant contributions of the present work. First, we confirmed that FEther maintains consistency between the Solidity source code and the respective formal specifications. To our knowledge, FEther is the first proof engine of Ethereum that supports the hybrid verification technology of Coq. Second, it provides a debug mechanism by which programmers can directly debug target smart contracts in Coq. Third, the correctness of FEther has been completely certified in Coq, implying that FEther is a reliable proof engine. Fourth, we provided a proprietary set of automatic tactics for FEther, which will help programmers to finish their property verifications with a high degree of automation. Finally, we optimized the high-level evaluation efficiency of FEther. We confirmed the utility of our previous works in building a certified executable proof engine in Coq.

### B. Extensibility and Universality

Obviously, the definitional interpreter of an intermediate must faithfully capture the intended behaviors of programs written in real-world programing languages. From a flexibility perspective, the same interpreter should also be applicable to multiple programing languages. Therefore, extensibility and universality were considered in the FEther design from the beginning of its development.

As mentioned in [14], we deliberately incorporated extensible space in Lolisa. This space is sufficient for expanding features such as pointer formalization and for implementing independent operator definitions. It can easily incorporate the features of mainstream programing languages by adding new typing rule constructors in the formal abstract syntax and the respective formal semantics. Moreover, the formal syntax of Lolisa is simplified by encapsulating it in syntax sugar notations $\mathcal{N}$. As shown in Rules 22 and 23, Lolisa is treated as the core formal language, which is transparent to real-world users. The formal syntax and semantics of Lolisa are logically classified into a general component $\mathcal{G}$ and $n$ special components $\mathcal{S}_i$ (see Rule 22 below). A general-purpose programming language $\mathcal{L}_i$ can be formalized identically to the Lolisa subset $\mathcal{G} \cup \mathcal{S}_i$, where $\mathcal{L}_i$ is symbolically represented by the syntax sugar notation $\mathcal{N}_i$. Here, the syntax symbols are nearly identical to the original syntax symbols of $\mathcal{L}_i$. This method assigns each $\mathcal{L}_i$ with a respective notation set $\mathcal{N}_i$ that satisfies $\mathcal{N}_i \subseteq Lolisa$. This relation, defined by Rule 23 below, also improves the extendibility of Lolisa.

$$Lolisa \stackrel{\text{def}}{=} \mathcal{G} \cup (\bigcup_{i=0}^{n} \mathcal{S}_i), \tag{22}$$
$$\forall i \in \mathbb{N}. \mathcal{L}_i \leftrightarrow \mathcal{N}_i \equiv \mathcal{G} \cup \mathcal{S}_i. \tag{23}$$

As the respective definitional interpreter of Lolisa, FEther inherits the extensibility advantages of Lolisa, and supports all of its syntaxes and semantics. Moreover, at the same level, any executable semantic $\mathcal{s}_i$ is independent of any other semantic, and all same-level semantics are encapsulated into an independent module $\mathcal{M}$ (see Rules 24 and 25 below). Higher-level semantics can access the APIs of lower-level semantics in different $\mathcal{M}$s, but the implementation details are transparent among the levels. Therefore, as shown in Figure 18, FEther is also easily extendible to new executable semantics in Lolisa without affecting the old semantics.

$$\mathcal{M}odule_K[\forall\, i, j \in \mathbb{N}, i \neq j.\, \mathcal{s}_i \cap \mathcal{s}_j = \emptyset], \quad (24)$$

$$\mathcal{s}_{h_m} \coloneqq \mathcal{M}_a.(I_i]\mathcal{s}_i]) \oplus \mathcal{M}_b.(I_j]\mathcal{s}_j]) \ldots \oplus \mathcal{M}_q.(I_n]\mathcal{s}_n]). \quad (25)$$

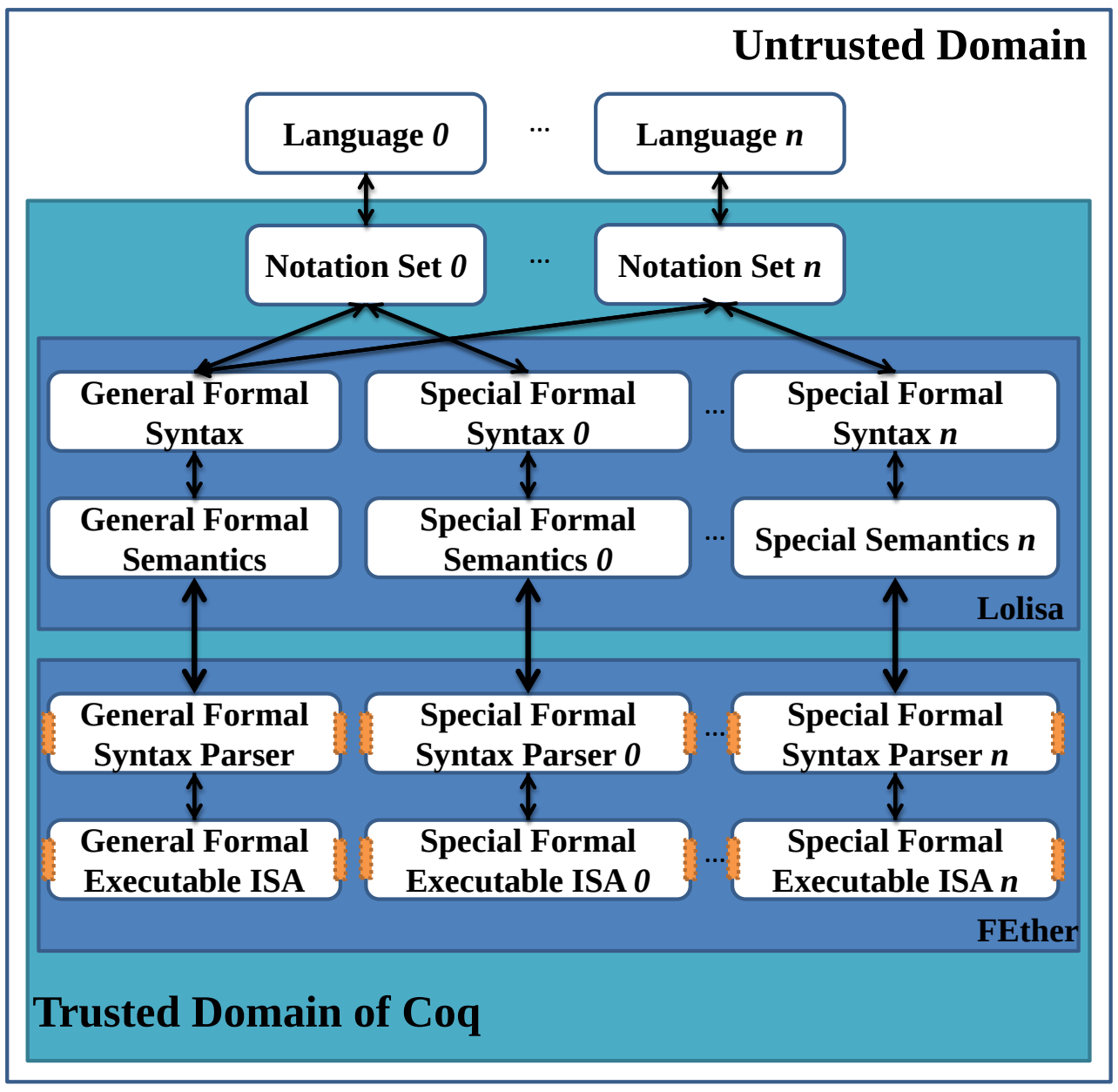

Figure 18. Detailed architecture for extending Lolisa to other general-purpose programing languages.

C. Limitations

Although the novel features in the current version of FEther confer many advantages, some limitations remain.

First, the FEther operates at the Solidity source-code level. Although it will not import vulnerabilities in the compiling process, it cannot guarantee the correctness of the bytecode when the compiler is untrusted. One possible solution is developing a low-level version of FEther that executes the bytecode generated by the compilation. One must then prove equivalence between the Solidity execution results and the respective execution results of the bytecode.

Second, similar to other symbolic execution tools, the FEther traverses all possible execution paths, which risks the path explosion problem. Given that Ethereum smart contracts are lightweight or even featherweight programs, however, the path explosion problem is almost precluded. Moreover, in situations that do meet the path explosion problem, the executions can be merged as invariants by the theorem-proving technology, and proven manually. This solution would exploit the selective symbolic execution pattern of FEther.

Finally, although the current version of FEther achieves property verifications by a few simple automatic tactics, it is not yet fully automated. In occasional situations, programmers must analyze the current proof goal and choose suitable verification tactics. Fortunately, this goal can be achieved by optimizing the design of the tactic evaluation strategies.

## 7. Conclusions and future work

This paper tackled the final challenge of the FSPVM blueprint: developing a definitional interpreter in Coq. The interpreter, called FEther, supports hybrid symbolic executions of Ethereum smart-contract formal verifications. Based on the GERM memory model, FEther accurately simulates the execution behaviors of Solidity in Coq. For evaluating complete situations during the FEther execution process, we also designed a set of tactics based on the Ltac mechanism of Coq, and combined them into a huge automatic tactic. With this tactic, the FEther can semi-automatically execute and verify different smart contracts in a symbolic virtual machine. To demonstrate the power of FEther in the real world, a sample smart contract was verified by conventional symbolic executions in FEther (simultaneous concolic and selective symbolic executions). We also compared the essential features of FEther and the cores of relevant tools. The self-correctness of FEther had been already confirmed by certifying the main functions in Coq. The current version of FEther supports the verification of smart contracts following the ERC20 standard. Finally, we discussed the extensibility and universality of FEther, and proposed an initial scheme for systematically simplifying and extending it, thus supporting the formalization of multiple general-purpose programming languages.

We hope that FSPVM-E will become sufficiently powerful and user-friendly for easy program verification by general programmers. Currently we are formalizing higher-level smart-contract development languages of the EOS blockchain platform [29]. We are also aiming to extend and optimize the current version of FEther. Future versions will support the assembly language of Solidity. Next, we will extend the FSPVM-E to support the Ethereum and EOS simultaneously. A formal verified interpreter of these languages will be developed based on the GERM platform. We will then build a general formal verification toolchain for blockchain smart contracts based on the EVI. Finally, we will build a general formal verification toolchain for blockchain smart contracts based on EVI, with the ultimate goal of automatic smart-contract verification.

**Acknowledgement**

The authors acknowledge the kind help of Marisa, and Enago for linguistic assistance during the preparation of this manuscript.

**Appendix A**

$$\mathcal{ESV}_{const} :: (\forall\ \tau_{const} : type, val\ \tau_{const}) \rightarrow Env \rightarrow Blc \rightarrow option\ value$$
$$\mathcal{ESV}_{const} \coloneqq \lambda\ (\ v_{const}(n) : (\forall\ \tau_{const} : type, val\ \tau_{const})).\lambda\ (env : Env).\lambda\ (b_{infor} : Blc).$$
$$Some\ m_{value_{const}}(n, env, b_{infor})$$

Table A.1 Semantics of constant Lolisa values

$$\mathcal{ESV}_{array} :: type \rightarrow index_{array} \rightarrow L_{address} \rightarrow Env \rightarrow memory \rightarrow Blc \rightarrow option\ value$$
$$\mathcal{ESV}_{array} \coloneqq$$
$$\lambda\ (\tau : type).\lambda\ (name : L_{address}).\lambda\ (env : Env).\lambda\ (\sigma : memory).\lambda\ (b_{infor} : Blc)\ .$$
$$\{|\ Some\ addr \longmapsto \left(\lambda\ (addr : L_{address})\ .read_{chck}(\ \sigma, env, b_{infor}, addr)\right)$$
$$Error \longmapsto Error\ |\}\ .\left(id_{search}(\tau, name, \sigma, env)\right)$$

Table A.2 Semantics of array types at the value layer

$$\mathcal{ESV}_{mapping} :: type \rightarrow type_{map} \rightarrow index_{map} \rightarrow option\ (val\ Tmap) \rightarrow Env \rightarrow memory \rightarrow Blc \rightarrow option\ value$$
$$\mathcal{ESV}_{mapping} \coloneqq$$
$$\lambda\ (\tau : type)\ .\lambda\ (\tau_{map} : type_{map})\ .\lambda\ (id : index_{map})\ .\lambda\ (env : Env).\lambda\ (\sigma : memory).\lambda\ (b_{infor} : Blc)\ .$$
$$\lambda\ \left(Vmap\left(name\ [id]\ (\tau_{map} \Rightarrow \tau)\ snd\right) : val\ (Tmap\ it\ t)\right).$$
$$\{|\ Some\ addr \longmapsto$$
$$\left(\lambda\ (addr : L_{address})\ .read_{chck}(\ \sigma, env, b_{infor}, addr)\right)\ addr$$
$$Error \longmapsto Error\ |\}\ .\left(id_{map}(id, env, \sigma, name, \tau, \tau_{map}, snd)\right)$$

Table A.3 Semantics of mapping values

$$\mathcal{ESV}_{str} :: L_{address} \rightarrow Env \rightarrow memory \rightarrow Blc \rightarrow option\ memory$$
$$\mathcal{ESV}_{str} \coloneqq$$
$$\lambda\ (str_v : L_{address}).\lambda\ (env : Env).\lambda\ (\sigma : memory).\lambda\ (b_{infor} : Blc).$$
$$read_{chck}(\ \sigma, env, b_{infor}, str_v)$$

Table A.4 Semantics of struct at the value layer

$$\mathcal{ESV}_{field} :: L_{address} \rightarrow list\ struct_{name} \rightarrow list\ ival \rightarrow Env \rightarrow memory \rightarrow Blc \rightarrow option\ value$$
$$\mathcal{ESV}_{field} \coloneqq$$
$$\lambda\ (head : L_{address}).\lambda\ (mems : list\ struct_{name}).\lambda\ (opars : list\ ival).\lambda\ (env : Env).\lambda\ (\sigma : memory).\lambda\ (b_{infor} : Blc).$$
$$\{|\ Some\ \ (a_{init}, a_{type}) \longmapsto$$
$$\{|\ Some\ (Dad, m_v) \longmapsto$$
$$\{|\ T_{Fid} \longmapsto\ \ set_{dad}(Dad, m_v, opars)$$
$$|\ F_{Fid} \longmapsto m_v\ |\}.Fid_{chck}(m_v)$$
$$|\ Error \longmapsto Error\ |\}.mems_{find}(\sigma, mems, env, b_{infor}, a_{init}, a_{type})$$
$$|\ Error \longmapsto Error\ |\}.eval_{head}(\sigma, env, b_{infor}, head)$$

Table A.5 Semantics of field access at the value layer

$$\mathcal{ESE}_{lexpr_{array}} :: type \rightarrow index_{array} \rightarrow L_{address} \rightarrow Env \rightarrow memory \rightarrow option\ L_{address}$$
$$\mathcal{ESE}_{lexpr_{array}} \coloneqq \lambda\ (\tau : type)\ .\ \lambda\ (name : L_{address}).\ \lambda\ (env : Env).\ \lambda\ (\sigma : memory)\ .\ id_{search}(\tau, name, \sigma, env)$$

Table A.6 Semantics of left array values at the expression layer

$$\mathcal{ESE}_{lexpr_{map}} :: index_{map} \rightarrow Env \rightarrow memory \rightarrow option\ L_{address}$$
$$\mathcal{ESE}_{lexpr_{map}} \coloneqq \lambda\ \left(id : index_{map}\right).\ \lambda\ (env : Env).\ \lambda\ (\sigma : memory).$$
$$\left(id_{map}\left(id, env, \sigma, name, \tau, \tau_{map}, snd\right)\right)$$

Table A.7. Semantics of field access at the value layer

$$\mathcal{ESE}_{lexpr_{const}} :: (\forall\ \tau : type, val\ \tau) \rightarrow Env \rightarrow memory \rightarrow option\ L_{address}$$
$$\mathcal{ESE}_{lexpr_{const}} \coloneqq \lambda\ \left(\ v : (\forall\ \tau : type, val\ \tau)\right).\ \lambda\ (env : Env).\ \lambda\ (\sigma : memory)$$
$$\{|\ Some\ Varray(index, \tau, name) \longmapsto \mathcal{ESE}_{lexpr_{array}}(Tarray\ index\ \tau, name, \sigma, env)$$
$$Some\ Vmap\left(name\ \left[id_{map}\right]\left(\tau_{map} \Rightarrow \tau\right) snd\right) \longmapsto \mathcal{ESE}_{lexpr_{map}}\left(id_{map}, env, \sigma, name, \tau, \tau_{map}, snd\right)$$
$$None\ \longmapsto None$$
$$|\}.\ lexpr_{check}(v)$$

Table A.8 Semantics of left constant values at the expression layer

$$\mathcal{ESE}_{lexpr_{eaddr'}} :: \left(\forall\ a : option\ L_{address} : expr\ eaddr(a)\ eaddr(a)\right) \rightarrow option\ L_{address}$$
$$\mathcal{ESE}_{lexpr_{eaddr'}} \coloneqq \lambda\ \left(Eaddr([\![name]\!]) : expr\ eaddr([\![name]\!])\ eaddr([\![name]\!])\right).\ [\![name]\!]$$

Table A.9 Semantics of reference expressions *Evar*, *Efun*, *Econ*, and *Epar* at the expression layer

$$\mathcal{ESE}_{rexpr_{const}} :: (\forall\ \tau : type, val\ \tau) \rightarrow Env \rightarrow memory \rightarrow Blc \rightarrow option\ value$$
$$\mathcal{ESE}_{rexpr_{const}} \coloneqq \lambda\left(v : (\forall\ \tau : type, val\ \tau)\right).\ \lambda\ (env : Env).\ \lambda\ (\sigma : memory).\ \lambda\ \left(b_{infor} : Blc\right).$$
$$\mathcal{ESV}_{value}(K, v, env, \sigma, b_{infor})$$

Table A.10 Semantics of right constant values at the expression layer

$$\mathcal{ESE}_{rexpr_{str}} :: \left(\forall\ a : L_{address}, struct_{par}\ a\right) \rightarrow Env \rightarrow memory \rightarrow Blc \rightarrow option\ value$$
$$\mathcal{ESE}_{rexpr_{str}} \coloneqq$$
$$\lambda\ (K : \mathbb{Z}).\ \lambda\ \left(head\ \{v_0; v_1; \ldots; v_n\} : struct_{par}\ head\right).\ \lambda\ (env : Env).\ \lambda\ (\sigma : memory).\ \lambda\ \left(b_{infor} : Blc\right)$$
$$eval_{str}\left(K, head\ \{v_0; v_1; \ldots; v_n\}, \sigma, env, b_{infor}\right)$$

Table A.11 Semantics of right struct values at the expression layer

$$\mathcal{ES}_{rexpr_{eaddr}} :: option\ L_{address} \rightarrow Blc \rightarrow Env \rightarrow memory \rightarrow option\ value$$
$$\mathcal{ES}_{rexpr_{eaddr}} \coloneqq \lambda\ (oaddr : option\ L_{address}).\ \lambda\ \left(b_{infor} : Blc\right).\ \lambda\ (env : Env).\ \lambda\ (\sigma : memory)\ .$$
$$\{|\ Some\ addr \longmapsto\ read_{chck}\left(\sigma, env, b_{infor}, addr\right)$$
$$None \longmapsto Error\ |\}\ .\ oaddr$$

Table A.12 Semantics of right reference values at the expression layer

$$\mathcal{ES}_{rexpr_{bop}} :: \forall(\tau\ \tau'\tau_1\ \tau_2 : type), bop_{\tau_1\ \tau_2} \rightarrow expr_{\tau\ \tau_1} \rightarrow expr_{\tau'\ \tau_1} \rightarrow Blc \rightarrow Env \rightarrow memory \rightarrow option\ value$$
$$\mathcal{ES}_{rexpr_{bop}} \equiv$$
$$\forall(\tau\ \tau'\tau_1\ \tau_2 : type), \lambda\left(op_2 : bop_{\tau_1\ \tau_2}\right).\lambda\left(e_0 : expr_{\tau\ \tau_1}\right).\lambda\left(e_1 : expr_{\tau'\ \tau_1}\right).\lambda\left(env : Env\right).\lambda\left(\sigma : memory\right).$$
$$eval_{bop}\left(\sigma, op_2, \mathcal{ESE}_{rexpr}(n, e_0, \sigma, blc, env), \mathcal{ESE}_{rexpr}(n, e_1, \sigma, blc, env)\right)$$

Table A.13 Semantics of right binary operations at the expression layer

$$\mathcal{ES}_{rexpr_{uop}} :: \forall(\tau_0\ \tau_1\ \tau_2 : type), uop_{\tau_1\ \tau_2} \rightarrow expr_{\tau_0\ \tau_1} \rightarrow Blc \rightarrow Env \rightarrow memory \rightarrow option\ value$$
$$\mathcal{ES}_{rexpr_{uop}} \coloneqq$$
$$\lambda\left(\tau_0\ \tau_1 : type\right).\lambda\left(op_1 : uop_{\tau_1\ \tau_2}\right).\lambda\left(e : expr_{\tau_0\ \tau_1}\right).\lambda\left(env : Env\right).\lambda\left(\sigma : memory\right).$$
$$eval_{uop}\left(\sigma, op_1, \mathcal{ESE}_{rexpr}(n, e, \sigma, blc, env)\right)$$

Table A.14 Semantics of right unary operations at the expression layer

$$\mathcal{ESS}_{con} :: option\ L_{address} \rightarrow list\ (prod\ value\ L_{address}) \rightarrow list\ L_{address} \rightarrow list\ L_{address} \rightarrow Env \rightarrow memory \rightarrow Blc \rightarrow option\ memory$$
$$\mathcal{ESS}_{con} \coloneqq \lambda\left(oname : option\ L_{address}\right).\lambda\left(con_{infor} : list\ (prod\ value\ L_{address})\right).\lambda\left(inherits, inhertis_c : list\ L_{address}\right).$$
$$\lambda\left(env : Env\right).\lambda\left(\sigma : memory\right).\lambda\left(b_{infor} : Blc\right).$$
$$\{|\ Some\ addr \longmapsto$$
$$\{|\ \ left\ _ \longmapsto Some\ write_{dir}\left(\left(\sigma, a, Cid\left((cid\ oaddr), con_{infor}, env, b_{infor}\right)\right)\right)$$
$$right\ _\ \longmapsto Some\ \sigma\ |\}.\left(inherit_{check}(inherits, inhertis_c)\right)$$
$$None \longmapsto None\ |\}.oname$$

Table A.15 Semantics of contract declarations

$$\mathcal{ESS}_{var} :: option\ L_{address} \rightarrow type \rightarrow option\ access \rightarrow Env \rightarrow memory \rightarrow Blc \rightarrow option\ memory$$
$$\mathcal{ESS}_{var} \coloneqq \lambda\left(oname : option\ L_{address}\right).\lambda\left(\tau : type\right).\lambda\left(oacc : option\ access\right).\lambda\left(env : Env\right).$$
$$\lambda\left(\sigma : memory\right).\lambda\left(b_{infor} : Blc\right).$$
$$\{|\ Some\ addr \longmapsto init_{var}(\sigma, env, b_{infor}, oacc, \tau, name)$$
$$None \longmapsto Error$$
$$|\}.oname$$

Table A.16 Semantics of variable declarations

$$\mathcal{ESS}_{str} :: L_{address} \rightarrow struct_{mem} \rightarrow option\ memory$$
$$\mathcal{ESS}_{str} \coloneqq \lambda\left(str_\tau : L_{address}\right).\lambda\left(mems_\tau : struct_{mem}\right).\lambda\left(env : Env\right).\lambda\left(\sigma : memory\right).\lambda\left(b_{infor} : Blc\right).$$
$$write_{dir}\left(\sigma, str_\tau, Str_{type}\left(str_\tau, mems_\tau, env, b_{infor}\right)\right)$$

Table A.17 Semantics of struct declarations

$$
\begin{aligned}
&\mathcal{ESS}_{call} :: \mathbb{Z} \to L_{address} \to option\ (list\ value) \to statement \to Env \to memory \to Blc \to option\ memory \\
&\mathcal{ESS}_{call} \coloneqq \lambda\ (pump : \mathbb{Z}).\ \lambda\ (name : L_{address}).\ \lambda\ \big(args : option\ (list\ value)\big).\ \lambda\ (s : statement). \\
&\qquad \lambda\ (env : Env).\ \lambda\ (\sigma : memory).\ \lambda\ \big(b_{infor} : Blc\big). \\
&\ \{|\ Some\ m_v \mapsto \\
&\quad \{|\ Some\ stt \mapsto \mathcal{ESS}\Big(pump, \sigma, args, set_{env}\Big(env, Fun_{call}\Big(\big(Efun(Some\ name)\big), args\Big)\Big), env, (stt ++ s)\Big) \\
&\qquad Error \mapsto None \\
&\quad |\}.\ get_{stt}(m_v) \quad set_{par}(\sigma', fpars, inputs) \\
&\quad Error \mapsto None \\
&\ |\}.\ read_{chck}\big(\sigma, env, b_{infor}, \alpha\big)
\end{aligned}
$$

Table A.18 Semantics of function call statements

$$
\begin{aligned}
&\mathcal{ESS}_{for} :: \mathbb{Z} \to (\forall\ \tau : type, expr_{\tau\ Tbool}) \to statement \to statement \to statement \to statement \\
&\qquad \to Env \to Env \to option\ (list\ value) \to memory \to Blc \to option\ memory \\
&\mathcal{ESS}_{for} \coloneqq \lambda\ (K : \mathbb{Z}).\ \lambda\ \big(e : (\forall\ \tau : type, expr_{\tau\ Tbool})\big).\ \lambda\ \big(s_0\ s_{body}\ s_1 : statement\big).\ \lambda\ (s_{next} : statement). \\
&\qquad \lambda\ (env : Env).\ \lambda\ (fenv : Env).\ \lambda\ \big(args : option\ (list\ value)\big).\ \lambda\ (\sigma : memory).\ \lambda\ \big(b_{infor} : Blc\big). \\
&\ \{|\ Some\ b \mapsto \\
&\quad \{|\ true \mapsto \\
&\qquad \{|\ Some\ \sigma' \mapsto \\
&\qquad\quad \{|\ Some\ \sigma'' \mapsto \\
&\qquad\qquad \{|\ Some\ \sigma''' \mapsto \mathcal{ESS}\big(K, \sigma, args, env, fenv, Loop_{for}\big(e, s_0, s_{body}, s_1\big) :: s_{next}\big) \\
&\qquad\qquad Error\ \mapsto Error\ |\}.\ \mathcal{ESS}(K, \sigma, args, env, fenv, s_1) \\
&\qquad\quad Error\ \mapsto Error\ |\}.\ \mathcal{ESS}\big(K, \sigma, args, env, fenv, s_{body}\big) \\
&\qquad Error\ \mapsto Error\ |\}.\ \mathcal{ESS}(K, \sigma, args, env, fenv, s_0) \\
&\quad false \mapsto \mathcal{ESS}(s_{next})\ |\}.\ b \\
&\ Error \mapsto Error\ |\}.\ get_{body}\Big(\mathcal{ESE}_{rexpr}(K, e, env, \sigma)\Big)
\end{aligned}
$$

Table A.19 Semantics of for loop statements

$$
\begin{aligned}
&\mathcal{ESS}_{while} :: \mathbb{Z} \to (\forall\ \tau : type, expr_{\tau\ Tbool}) \to statement \to statement \to Env \to Env \\
&\qquad \to option\ (list\ value) \to memory \to Blc \to option\ memory \\
&\mathcal{ESS}_{while} \coloneqq \lambda\big(e : (\forall\ \tau : type, expr_{\tau\ Tbool})\big).\ \lambda\ \big(s_{body} : statement\big).\ \lambda\ (s_{next} : statement). \\
&\qquad \lambda\ (env : Env).\ \lambda\ (fenv : Env).\ \lambda\ \big(args : option\ (list\ value)\big).\ \lambda\ (\sigma : memory).\ \lambda\ \big(b_{infor} : Blc\big). \\
&\ \{|\ Some\ b \mapsto \\
&\quad \{|\ true \mapsto \\
&\qquad \{|\ Some\ \sigma' \mapsto\ \mathcal{ESS}\big(K, \sigma, args, env, fenv, Loop_{while}\big(e, s_{body}\big) :: s_{next}, \sigma'\big) \\
&\qquad Error\ \mapsto Error\ |\}.\ \mathcal{ESS}\big(K, \sigma, args, env, fenv, s_{body}\big) \\
&\quad false \mapsto \mathcal{ESS}(s_{next})\ |\}.\ b \\
&\ Error \mapsto Error\ |\}.\ get_{bool}\Big(\mathcal{ESE}_{rexpr}(K, e, env, \sigma)\Big)
\end{aligned}
$$

Table 20. Semantics of while loop statements

$$\mathcal{ESS}_{modif} :: statement \rightarrow Env \rightarrow Env \rightarrow memory \rightarrow Blc \rightarrow option\ memory$$

$$\mathcal{ESS}_{modif} \coloneqq \lambda\ (s : statement).\ \lambda\ (env : Env).\ \lambda\ (fenv : Env).\ \lambda\ (\sigma : memory).\ \lambda\ (b_{infor} : Blc).$$

$$\{|\ Some\ fun_{infor}(Efun(Tbool, (Some\ a)), [fpar_0, fpar_1, \dots, fpar_n], \Lambda_{fun}, s_{body}) \mapsto$$

$$\{|\ Some\ \sigma' \mapsto$$

$$\{|\ Some\ \sigma'' \mapsto write_{chck}(\sigma'', env, b_{infor}, name, s_{body})$$

$$None\ \mapsto None\ |\}.\ init_{re}(\sigma', \Lambda_{fun}, [Tbool])$$

$$None\ \mapsto None\ |\}.\ repeat\ (init_{var}(\sigma, env, fenv, oacc, \tau, fpar_i), [fpar_0, fpar_1, \dots, fpar_n])$$

$$None\ \mapsto None\ |\}.\ get_{fun}(s)$$

Table A.21 Semantics of modifier statements

$$\mathcal{ESS}_{fun} :: statement \rightarrow Env \rightarrow Env \rightarrow memory \rightarrow Blc \rightarrow option\ memory$$

$$\mathcal{ESS}_{fun} \coloneqq \lambda\ (s : statement).\ \lambda\ (env : Env).\ \lambda\ (fenv : Env).\ \lambda\ (\sigma : memory).\ \lambda\ (b_{infor} : Blc).$$

$$\{|\ Some\ [\sigma_0, \sigma_1, \dots, \sigma_n] \mapsto$$

$$\{|\ Some\ true \mapsto$$

$$\{|\ Some\ fun_{infor}(Efun([\tau_0, \tau_1, \dots, \tau_n], (Some\ a)), [fpar_0, fpar_1, \dots, fpar_n], \Lambda_{fun}, s_{body}) \mapsto$$

$$\{|\ Some\ \sigma' \mapsto$$

$$\{|\ Some\ \sigma'' \mapsto write_{chck}(\sigma'', env, b_{infor}, name, s_{body})$$

$$None\ \mapsto None\ |\}.\ init_{re}(\sigma', \Lambda_{fun}, [\tau_0, \tau_1, \dots, \tau_n])$$

$$None\ \mapsto None\ |\}.\ repeat\ (init_{var}(\sigma, env, fenv, oacc, \tau, fpar_i), [fpar_0, fpar_1, \dots, fpar_n])$$

$$None\ \mapsto None\ |\}.\ get_{fun}(s)$$

$$Some\ false\ \mapsto \sigma$$

$$None\ \mapsto None\ |\}.\ modif_{chck}([\sigma_0, \sigma_1, \dots, \sigma_n], \sigma_{mtrue}, \sigma)$$

$$None\ \mapsto None\ |\}.\ repeat\ (\mathcal{ESS}_{call}(\sigma, modi_i), [modi_0, modi_1, \dots, modi_n])$$

Table A.22 Semantics of function statements

$$\mathcal{ESS}_{if} :: (\forall\ \tau : type, expr_{\tau\ Tbool}) \rightarrow statement \rightarrow statement \rightarrow Env \rightarrow memory \rightarrow Blc \rightarrow option\ statement$$

$$\mathcal{ESS}_{if} \coloneqq \lambda(e : (\forall\ \tau : type, expr_{\tau\ Tbool})).\ \lambda\ (s_0\ s_1 : statement).\ \lambda\ (\sigma : memory).\ \lambda\ (b_{infor} : Blc).$$

$$\{|\ Some\ b \mapsto$$

$$\{|\ true \mapsto Some\ s_0$$

$$false \mapsto Some\ s_1\ |\}.\ b$$

$$Error \mapsto Error\ |\}.\ get_{bool}\left(\mathcal{ESE}_{rexpr}(K, e, env, \sigma)\right)$$

Table A.23 Semantics of conditional statements

$$\mathcal{ESS}_{assign} :: \mathbb{Z} \to (\forall\, \tau_0, \tau_1 : type, expr_{\tau_0\, \tau_1}) \to (\forall\, \tau_0, \tau_1 : type, expr_{\tau_0\, \tau_1}) \to statement \to statement \to option\ statement$$
$$\mathcal{ESS}_{assign} \coloneqq \lambda\,(K : \mathbb{Z}).\, \lambda\,(e_r : (\forall\, \tau : type, expr_{\tau\, Tbool})).\, \lambda\,(s_0\, s_1 : statement).\, \lambda\,(\sigma : memory).\, \lambda\,(b_{infor} : Blc).$$
$$\{|\ Some\ v \longmapsto$$
$$\{|\ Some\ name \longmapsto$$
$$\{|\ Some\ v \longmapsto$$
$$\{|\ true \longmapsto \mathcal{ESS}\left(K, \sigma, args, env, fenv, Fun_{call}\left((Efun(o\alpha)), inputs\right)\right)$$
$$false \longmapsto write_{chck}(\sigma', env', b_{infor}, addr, v')$$
$$|\}.\, check_v(Fid, v)$$
$$Error \longmapsto Error$$
$$|\}.\, \mathcal{ESE}_{lexpr}(K, e, env, \sigma)$$
$$Error \longmapsto Error$$
$$|\}.\, \mathcal{ESE}_{rexpr}(K, e, env, \sigma)$$

Table A.24 Semantics of assignment statements

$$\mathcal{ESS}_{re} :: (\forall\, \tau_0\, \tau_1 : type, expr_{\tau_0\, \tau_1}) \to memory \to Env \to Env \to Blc \to option\ memory$$
$$\mathcal{ESS}_{re} \coloneqq \lambda\,(\forall\, \tau_0\, \tau_1 : type, expr_{\tau_0\, \tau_1}).\, \lambda\,(\sigma : memory).\, \lambda\,(env\ fenv : Env).\, \lambda\,(b_{infor} : Blc).$$
$$\{|\ Some\ v \longmapsto write_{chck}(\sigma, env, b_{infor}, \Lambda_{fun}, v)$$
$$Error \longmapsto Error\ |\}.\, \mathcal{ESE}_{rexpr}(K, e, env, \sigma)$$

Table A.25 Semantics of return statements

$$\mathcal{ESS}_{res} :: exprs \to memory \to Env \to Env \to Blc \to option\ memory$$
$$\mathcal{ESS}_{res} \coloneqq \lambda\,(es : exprs).\, \lambda\,(env : Env).\, \lambda\,(b_{infor} : Blc).\, \lambda\,(\sigma : memory)$$
$$\{|\ Some\ vs \longmapsto write_{chck}(\sigma, env, b_{infor}, \Lambda_{fun}, vs)$$
$$Error \longmapsto Error\ |\}.\, eval_{exprs}(K, es, env, \sigma)$$

Table A.26 Semantics of multiple return statements

$$\mathcal{ESS}_{snil} :: memory \to option\ memory$$
$$\mathcal{ESS}_{snil} \coloneqq \lambda\,(\sigma : memory).\,(Some\ \sigma)$$
$$\mathcal{ESS}_{throw} :: memory \to option\ memory$$
$$\mathcal{ESS}_{throw} \coloneqq \lambda\,(\sigma : memory).\,\left(Some\ write_{dir}(\sigma_{init}, env, b_{infor}, _0xthrow, true)\right)$$
$$\mathcal{ESS}_{funstop} :: memory \to option\ memory$$
$$\mathcal{ESS}_{funstop} \coloneqq \lambda\,(\sigma : memory).\,(Some\ \sigma)$$

Table A.27 Semantics of skip, throw and function stop statements

**Appendix B**

Algorithm B.1. Source code of the *wallet* function

```
Function wallet () public payable {
    uint index = indexes[msg.sender];
    uint open; uint close; uint quota; uint rate; uint partiLimit;
    uint totalLimit; uint finalLimit;

    if (privileges[msg.sender]) {
      open = privilegeOpen;
      close = privilegeClose;
      quota = privilegeQuota;
      rate = RATE_PRIVILEGE;
    } else {
      open = ordinaryOpen;
      close = ordinaryClose;
      quota = ordinaryQuota;
      rate = RATE_ORDINARY;
    }

    if (now < open || now > close) {
      revert();
    }
    if (subscription >= TOKEN_TARGET_AMOUNT) {
      revert();
    }
    if (index == 0) {
      revert();
    }
    if (deposits[index] >= quota) {
      revert();
    }
    if (msg.value == 0) {
      revert();
    }
    if (msg.value % 1000000000000000000 != 0) {
      revert();
    }
    partiLimit = quota - deposits[index];
    totalLimit = ((TOKEN_TARGET_AMOUNT - subscription)
     - (TOKEN_TARGET_AMOUNT - subscription) % rate) / rate *
1000000000000000000;

    if (partiLimit <= totalLimit) {
      finalLimit = partiLimit;
    } else {
      finalLimit = totalLimit;
    }

    if (msg.value <= finalLimit) {
      safe.transfer(msg.value);
      deposits[index] += msg.value;
      subscription += msg.value / 1000000000000000000 * rate;
      Transfer(msg.sender, msg.value);
    } else {
      safe.transfer(finalLimit);
      deposits[index] += finalLimit;
      subscription += finalLimit / 1000000000000000000 * rate;
      Transfer(msg.sender, finalLimit);
      msg.sender.transfer(msg.value - finalLimit);
    }
}
```